\documentclass[a4paper,11pt]{article}
\pdfoutput=1 

\usepackage{jcappub} 

\usepackage[varg]{txfonts}
\usepackage[utf8]{inputenc}

\usepackage{longtable}
\usepackage{epsfig,caption}
\usepackage{color}

\usepackage{lineno} 

\usepackage[toc,page]{appendix}

\newcommand\ion[2]{#1$\,${\scshape{#2}}}
\newcommand{\lya}{Ly\ensuremath{\alpha}}%

\title{\bf  The one-dimensional power spectrum from the SDSS DR14 Ly$\alpha$ forests }

\author[a]{Sol\`ene Chabanier,}
\author[a]{Nathalie Palanque-Delabrouille,}
\author[a]{Christophe Y\`eche,}
\author[a]{Jean-Marc Le Goff,}
\author[a]{Eric Armengaud,}
\author[b]{Julian Bautista,}
\author[c]{Michael Blomqvist,}
\author[d]{Nicolas Busca,}
\author[e]{Kyle Dawson,}
\author[a]{Thomas Etourneau,}
\author[f]{Andreu Font-Ribera,}
\author[g]{Youngbae Lee,}
\author[e]{H\'elion du Mas des Bourboux,}
\author[c]{Matthew Pieri,}
\author[a]{James Rich,}
\author[g]{Graziano Rossi,}
\author[h]{Donald Schneider,}
\author[i]{An\v{z}e Slosar}

\emailAdd{solene.chabanier@cea.fr, nathalie.palanque-delabrouille@cea.fr, christophe.yeche@cea.fr, jmlegoff@cea.fr}

\affiliation[a]{IRFU, CEA, Université Paris-Saclay, F-91191 Gif-sur-Yvette, France}
\affiliation[b]{Institute of Cosmology \& Gravitation, University of Portsmouth, Dennis Sciama Building, Portsmouth, PO1 3FX, UK}
\affiliation[c]{Aix Marseille Univ, CNRS, LAM, Laboratoire d'Astrophysique de Marseille, Marseille, France}
\affiliation[d]{Sorbonne Universit\'e, Universit\'e Paris Diderot, CNRS/IN2P3, Laboratoire de Physique Nucl\'eaire et de Hautes Energies, LPNHE, 4 Place Jussieu, F-75252 Paris, France}
\affiliation[e]{Department of Physics and Astronomy, University of Utah, 115 S 1400 E, Salt Lake City, UT 84112, USA}
\affiliation[f]{Department of Physics and Astronomy, University College London, Gower Street, London, United Kingdom}
\affiliation[g]{Department of Physics and Astronomy, Sejong University, Seoul, 143-747, Korea}
\affiliation[h]{Department of Astronomy and Astrophysics, The Pennsylvania State University,
   University Park, PA 16802, USA \\  Institute for Gravitation and the Cosmos, The Pennsylvania State University,
   University Park, PA 16802, USA}
\affiliation[i]{Physiscs Department, Brookhaven National Laboratory, Upton NY 11973}
\date{Received xx; accepted xx}

\abstract{ 
We present a measurement of the 1D Ly$\alpha$ forest flux power spectrum, using  the complete Baryon Oscillation Spectroscopic Survey (BOSS)  and  first  extended-BOSS (eBOSS) quasars at $z_{\rm qso}>2.1$, corresponding to the fourteenth data release (DR14)  of the Sloan Digital Sky Survey (SDSS). Our results cover thirteen bins in redshift from $z_{\rm Ly\alpha}=2.2$ to 4.6, and scales up to $k=0.02\rm \,(km/s)^{-1}$. From a parent sample of 180,413 visually inspected spectra, we selected the 43,751 quasars with the best quality; this data set improves  the previous  result from  the ninth data release (DR9), both in statistical precision (achieving a reduction by a factor of two) and in redshift coverage. We also present a thorough investigation of  identified sources of systematic uncertainties that  affect the measurement. The resulting 1D power spectrum of this work is in excellent agreement with the one from the BOSS DR9 data. 
}

\begin{document}
\maketitle
\flushbottom

\section{Introduction}
\label{sec:intro}
The neutral hydrogen present in the intergalactic medium imprints a characteristic absorption  in the spectra of distant quasars, known as the Lyman-$\alpha$ (Ly$\alpha$) forest~\cite{Gunn1965,Lynds1971}. This absorption of the quasar transmitted flux fraction caused by these absorptions can be accurately described by the flux power spectrum, or equivalently by the  flux auto-correlation, which has  been shown for over a decade to be a powerful tool in astrophysics and cosmology on scales ranging from a few to several hundred Mpc. 
On scales of a few Mpc, the Ly$\alpha$ flux power spectrum is an excellent probe of the thermal properties of the photo-ionized warm intergalactic medium (IGM)~\cite{Zaldarriaga2001,Meiksin2009, McQuinn2016}. The power spectrum can be used to set constraints on its temperature at various epochs~\cite{Viel2005, Bolton2008, Bolton2014, Garzilli2012, Garzilli2015}. On these small scales, it can also be used to measure the clustering properties of structures in the Universe at redshift 2 to 5, since it is sensitive to the smoothing caused by the free-streaming of relativistic particles. 
Combined with dedicated hydrodynamic simulations~\cite{Borde2014,Rossi2014,Bolton2017}, the Ly$\alpha$ flux power spectrum can  allow one to set constraints on the mass of neutrinos~\cite{Seljak2005, Viel2010, Palanque-Delabrouille2015, Palanque2015a, Yeche2017}, on dark radiation~\cite{Rossi2015}, on the possible existence of sterile neutrinos~\cite{Baur2016,Baur2017} or on the nature of dark matter~\cite{Armengaud2017,Irsic2017,Irsic2017b}. On scales of tens to hundreds of Mpc, the Ly$\alpha$ forest becomes a unique probe of the dark matter and dark energy content of the Universe at redshift $z\sim 2.5$, whether through the auto-correlation of the Ly$\alpha$ forest~\cite{Slosar2013, Busca2012,  Delubac2015, Bautista2017}, or through its cross-correlation with other matter tracers~\cite{Font-Ribera2014, DuMasDesBourboux2017}. 

While Ly$\alpha$ physics has long been studied with only a handful of  quasar spectra~\cite{McDonald2000,Croft2002, Kim2004}, there are now numerous  data sets that can be split into two main categories. The first group are the few hundred high-resolution spectra from, e.g., Keck/HIRES, VLT/UVES, or X-Shooter data~\cite{Viel2004, Viel2008, Viel2013, Irsic2016, Yeche2017, Walther2018}. They are typically used to constrain the thermal state of the IGM, but lack the large-scale modes that are needed to constrain cosmology. The second set are the thousand to several hundred  thousand medium-resolution spectra~\cite{McDonald2006, Palanque-Delabrouille2013} from the first iteration of the Sloan Digital Sky Survey (SDSS-I)~\cite{York2000} or  its later component the Baryon Oscillation Spectroscopic Survey (BOSS)~\cite{Dawson2013} of SDSS-III~\cite{Eisenstein2011}, from which most of the cosmological results quoted above were derived. 
The SDSS data are  sometimes used in combination with mid and high-resolution data to constrain both large and small scales simultaneously, as well as to improve the redshift level-arm that the data cover.  

In this work, we focus on  data measured with  BOSS and  extended BOSS (eBOSS)~\cite{Dawson2016} of the SDSS-III and SDSS-IV~\cite{Blanton2017} surveys, respectively. We compute the 1D flux power spectrum in the Ly$\alpha$ forest from a parent sample of 180,413 visually inspected eBOSS quasar spectra, a more than  three-fold  improvement over the latest SDSS measurement that was obtained with the first year BOSS data only~\cite{Palanque-Delabrouille2013}. We also pay close attention to the systematic uncertainty budget, which the present analysis  improves  compared to  previous publications. This work builds upon our previous analysis presented in~\cite{Palanque-Delabrouille2013}. To make it easier for the reader to identify any reference to this earlier paper, we will henceforth refer to it as PYB13.

The outline of the paper is as follows. Sec.~\ref{sec:Pk} presents the observational data and the value-added catalogs, and explains the general scheme for the measurement of the power spectrum. The  analysis of the data and the various ingredients that enter the computation of the power spectrum are detailed in Sec.~\ref{sec:analysis}. Sec.~\ref{sec:mocks} describes the dedicated synthetic data (thereafter {\em mocks})  and  shows how they are used to quantify and correct for the possible biases introduced in the analysis. We study the sources of systematic uncertainties in Sec.~\ref{sec:systs}. Finally, Sec.~\ref{sec:results} presents  the measured 1D flux power spectrum and  the resulting improvements  on cosmological constraints. We conclude in Sec.~\ref{sec:conclusion}.


\section{Power spectrum estimation}
\label{sec:Pk}
\subsection{SDSS data}

The results presented here are based on data collected by the SDSS~\cite{York2000}. We select our sample of \lya\ forest observations from the quasar spectra of the DR14Q catalog~\cite{Paris2018}, which were observed either over a five-year period from 2009 to 2014 by the SDSS-III Collaboration~\cite{Gunn2006,Ahn2012, Dawson2013, Eisenstein2011,Smee2013} during the BOSS survey, or in 2014 -- 2015 by the SDSS-IV Collaboration~\cite{Blanton2017} as part of the  eBOSS survey~\cite{Dawson2016}. The selection of the quasars for either survey is extensively described in~\cite{Ross2012,Myers2015,Palanque2016}. 

While all the quasar targets of the BOSS DR12Q quasar catalog~\cite{Paris2012} underwent visual inspection of the measured spectra, this is no longer the case for eBOSS. This is due to the significant increase in the number of targets, from $40~\rm deg^{-2}$ quasar targets  focusing on redshifts above 2.1 for BOSS, to about $115~\rm deg^{-2}$ quasar targets for eBOSS, encompassing both quasars in the redshift range $0.8<z_{\rm qso}<2.2$ used as direct matter tracers~\cite{Ata2018}, and quasars at $z_{\rm qso}>2.1$ used for \lya\ studies. As a consequence, there is a slight increase in the rate of inaccurate redshift determinations, which mostly affects quasars at redshifts above $\sim 4$ when the \ion{Mg}{ii} emission  line of a low-redshift quasar is mistaken for the \lya\ emission  line of a high-redshift quasar. This contamination, which is less than $1\%$,  has negligible impact on Baryon Acoustic Oscillation (BAO) studies where the bulk of the sample is largely dominated by $z_{\rm qso}\sim2.5$ quasars, but it is significant for 1D power spectrum analyses where all redshift bins  are considered with equal importance. In particular, for $z_{\rm qso}\gtrsim 4$, the fraction of quasars with a wrong assignation of the redshift  can reach  $30\%$ of the sample. In this work, we therefore restrict the sample to quasars observed before MJD=56870 when they were all visually inspected. We  use the latest spectral identification, i.e. the one given in DR14Q. 

The data are processed with release v5\_7\_0 based on the standard DR12 SDSS-III pipeline~\cite{Bolton2012}.
We use  the ``coadded" spectra constructed from typically four exposures of 15 minutes resampled at wavelength pixels of width $\Delta \log_{10}\lambda=10^{-4}$, or equivalently $\Delta v = c \Delta\lambda/\lambda = c\Delta \ln\lambda = 69\,{\rm (km/s)^{-1}}$. We also use the individual exposures to improve the estimate of  the pixel noise. The noise estimate procedure is described in Sec.~\ref{sec:noise}. 

\subsection{Value added catalogs} \label{sec:vac}

In addition to the SDSS quasar catalogs, we make use of additional catalogs and information which help in the selection of the spectra for the analysis. 
For instance, we want to exclude spectra exhibiting Damped \lya\ Absorptions (DLAs)  or quasars affected by broad absorption lines (BAL), since these features are not included in the simulations we use to compute cosmological constraints. 
We  use the Balnicity Index available in the DR14Q catalog, which flags quasars with BALs  in their spectra. Finally, we examine two external catalogs to identify regions in the quasar spectra affected by DLA systems. The first catalog is an update on DR14Q of the identification of DLAs following the fully automated procedure  described in~\cite{Noterdaeme2012}. We will hereafter refer to it as N12. The second catalog is constructed from a convolutional neural network designed to identify strong neutral H absorptions~\citep{Parks2018}. We will refer to the latter catalog as P18. Details on the use of these additional catalogs for the forest selection are given in Sec.~\ref{sec:selection}, and for the estimate of the associated systematic uncertainties  in Sec.~\ref{sec:mocks}. We also use a list of wavelength regions contaminated by sky emission lines. This list is available at \texttt {https://github.com/igmhub/picca/blob/master/etc/list\_veto\_line\_Pk1D.txt}

\subsection{\lya\ forest: definition and transmitted flux} \label{sec:transmittedflux}

\begin{figure}[htbp]
\begin{center}
\epsfig{figure= 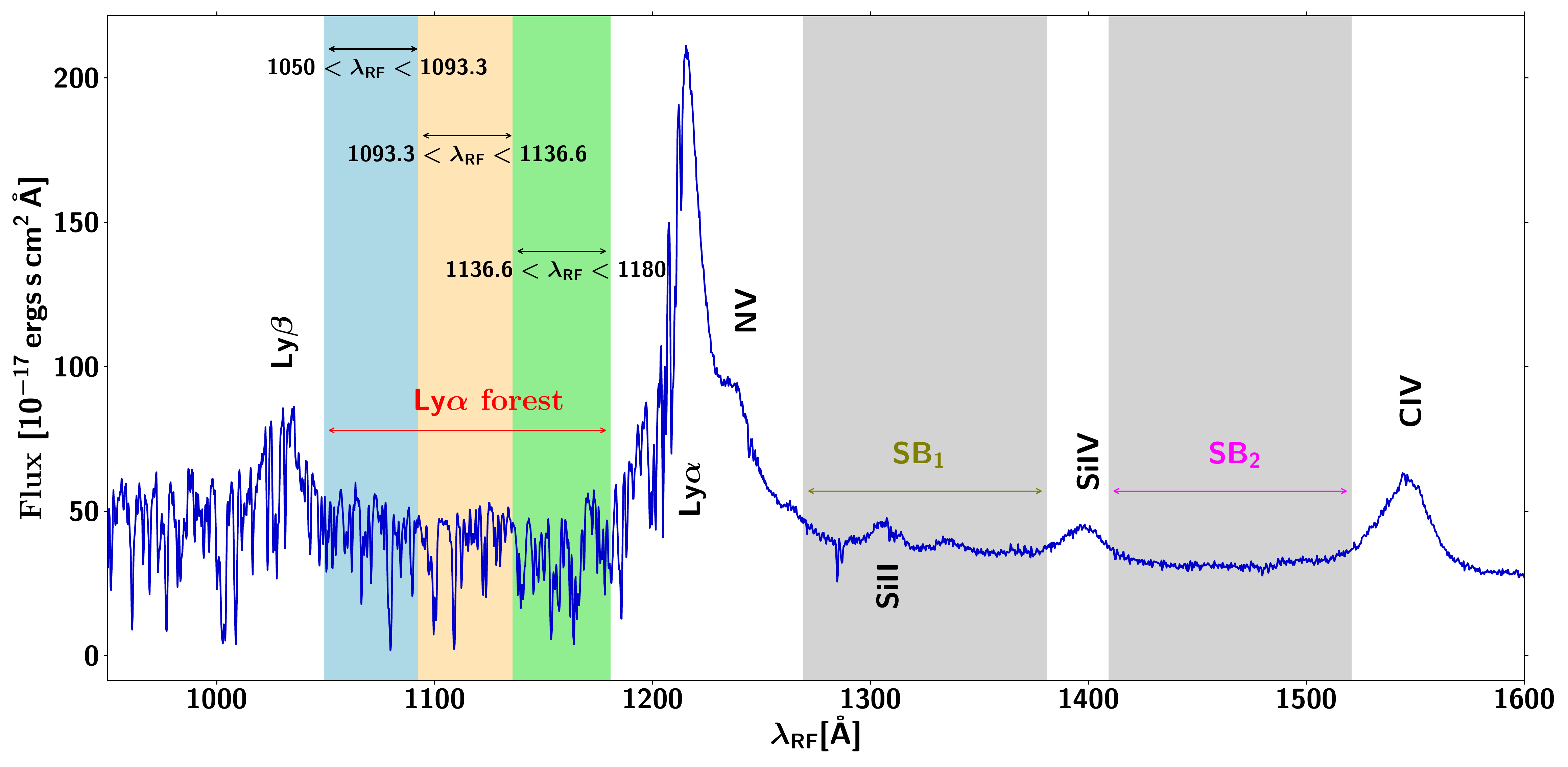,width =  \textwidth}\\
\epsfig{figure= 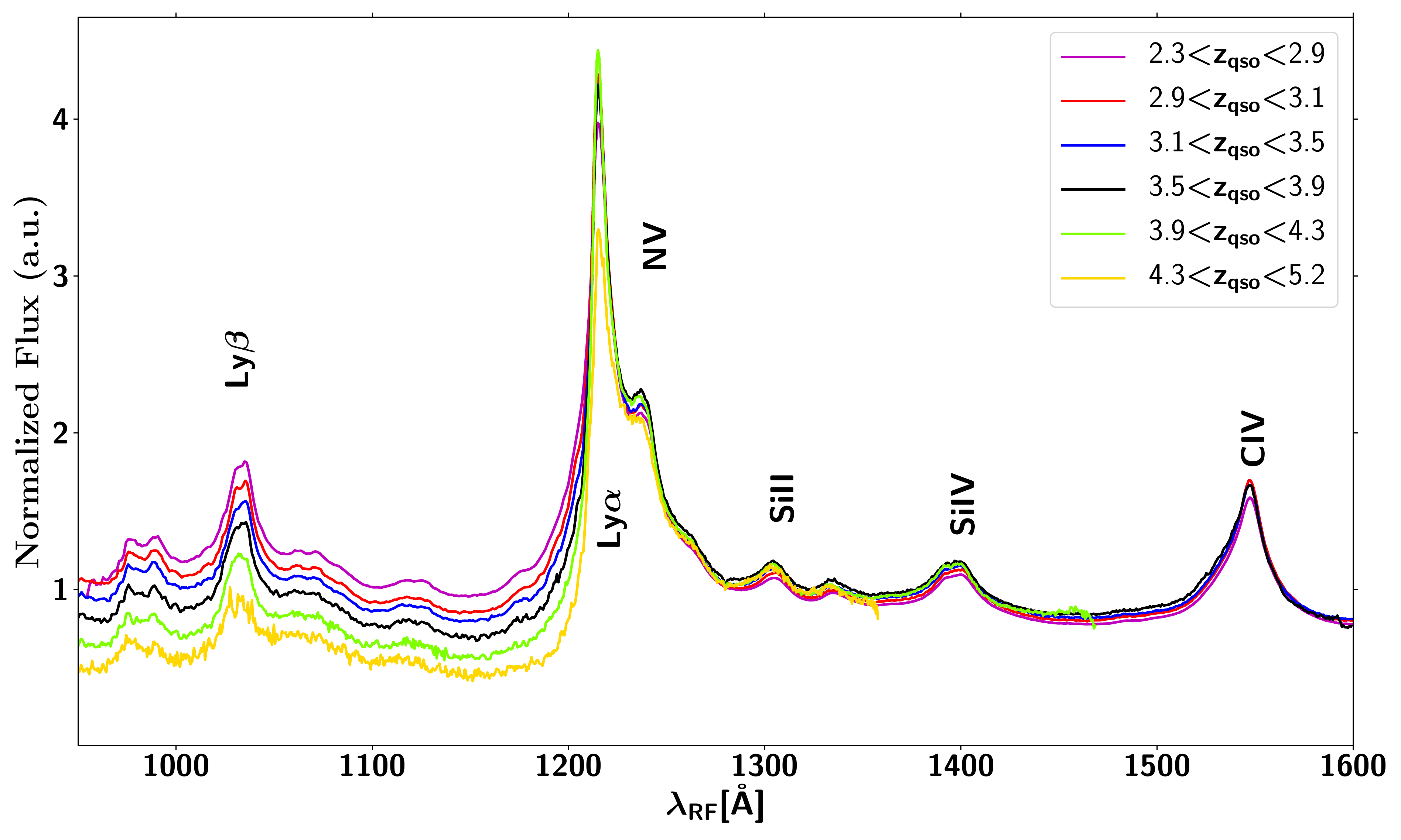,width =  \textwidth}
\caption{\it Top: Example of a bright quasar spectrum (SDSS  J114308.87+345222.2)  at redshift  $z_{qso} = 3.155$, observed by BOSS.  From the left to the right, the three colored bands represent the Ly$\alpha$ forest region ($1050<\lambda_{\rm RF}<1180\,\AA$), while the grey bands show  the first (${\rm SB_1}$ at $1270 < \lambda_{\rm RF}< 1380 \,\AA$)  and the second  (${\rm SB_2}$ at $1410 < \lambda_{\rm RF}< 1520 \,\AA$) side bands, in the quasar rest frame. Bottom:  Composite quasar spectra in six redshift bins from 2.3 to 5.2. All spectra are normalized at $\lambda_{RF} = 1280$\,\AA. } 
\label{fig:Spectra_eBOSS}
\end{center}
\end{figure}

The top plot of Fig.~\ref{fig:Spectra_eBOSS} displays one of our eBOSS spectra. The broad quasar emission lines are clearly visible, such as Ly$\beta$  ($1025.72\,$\AA), Ly$\alpha$  ($1215.67\,$\AA), \ion{N}{v}   ($1238.82 / 1242.80\,$\AA), \ion{Si}{iv}   ($1393.76/$ $1402.77\,$\AA) and \ion{C}{iv}  ($1548.20 / 1550.78\,$\AA), with all wavelengths  expressed in rest frame.  Ly$\alpha$ absorpsion along the quasar line of sight, constituting the \lya\ forest, appears bluewards of the quasar Ly$\alpha$ emission peak. For illustration purposes, the bottom panel of Fig.~\ref{fig:Spectra_eBOSS} shows  composite  spectra obtained by averaging  all 43,751 eBOSS quasar spectra  used in this analysis, split into six redshift bins. We can clearly see the higher mean absorption  (and hence smaller transmitted flux) at higher redshift, due to the larger density of neutral hydrogen as one moves to earlier times.

We define the Ly$\alpha$ forest region by the range $1050<\lambda_{\rm RF}<1180\,$\AA\ (colored bands in  Fig.~\ref{fig:Spectra_eBOSS}, top panel),   about $6000\,{\rm km/s}$ and $8500\,{\rm km/s}$    from the quasar Ly$\beta$ and Ly$\alpha$ emission peaks, respectively, to avoid contamination of the power spectrum by astrophysical effects in the vicinity of the quasar. The Ly$\alpha$ forest region spans a  redshift range $\Delta z \sim 0.4$ for a quasar at a redshift $z_{\rm qso}=3$, and $\Delta z \sim 0.6$ at $z_{\rm qso}=4.6$.  In order to improve the redshift resolution of the measured power spectrum and to reduce the correlation between redshift bins,  we split this range into  three  consecutive and non-overlapping sub-regions of  equal  length, each covering $\sim 170$  pixels of eBOSS spectra. The boundaries between these sub-regions are set at rest-frame wavelengths of 1093.3 and 1136.6~\AA. For the sake of simplicity, we hereafter use {\em forest} (and not {\em a third of forest} or {\em a sub-forest}, for instance) to refer to  each of these sub-regions. Each forest spans at most $\Delta z=0.2$.  The first step of the analysis is to extract these forests from the spectra; All subsequent steps  are applied on  each forest. 

The  transmitted flux  fraction $\delta(\lambda)$ in a forest is estimated from the pixel flux $f(\lambda)$ by
\begin{equation}
\delta(\lambda) = \frac{f(\lambda)} {C_q(\lambda) \overline{F}(z_{\rm Ly\alpha})} - 1\, ,
\label{eq:delta}
\end{equation}
where  $C_q(\lambda)$ is the  unabsorbed flux of a quasar (the mean quasar `continuum') and $\overline{F}(z_{\rm Ly\alpha})$ is the mean transmitted flux fraction at the \ion{H}{i} absorber redshift.  For a pixel at observed wavelength $\lambda$, the corresponding \ion{H}{i} absorber redshift $z_{\rm Ly\alpha}$ can be inferred from $1+z_{\rm Ly\alpha} = \lambda / \lambda_{\rm Ly\alpha}$, where $\lambda_{\rm Ly\alpha} = 1215.67\,$\AA. The product $ C_q(\lambda) \overline{F}(z_{\rm Ly\alpha})$  is determined using a method  similar to the approach developed in~\cite{Bautista2017,DuMasDesBourboux2017}. 
We assume that the quasar continuum  is the product of a universal function of the rest-frame wavelength, $\lambda_{RF}=\lambda/(1+z_{\rm qso})$, and  a quasar-dependent factor $a_q$:
\begin{equation}
C_q(\lambda)=a_q C(\lambda_{RF})\; ,
\label{eq:continuum}
\end{equation}
where $C(\lambda_{RF})$ assumes no functional form, and is normalized so that its integral over the forest is equal to unity. The $a_q$ and $C(\lambda_{RF})$ are determined  iteratively by maximizing the likelihood function 
\begin{equation}
{\cal L} = \prod_{q,\lambda} P(\, f(\lambda)\;|\;C_q(\lambda)\,) \;.
\end{equation} 
Here $P(f(\lambda)\;|\;C_q(\lambda))$ is the probability to observe a flux $f(\lambda)$
for a given continuum found by convolving the intrinsic probability (model described in~\cite{Bautista2015})
with the observational resolution assumed to be Gaussian. 

The method uses the same publicly available pipeline \texttt{picca}\footnote{\tt https://github.com/igmhub/picca} (Package for IGM Cosmological-Correlations Analyses). 
In contrast to BAO analyses where pixels are weighted by the noise inverse variance, we here do not any apply  weights. All  pixels of  a given quasar thus contribute equally to the measurement of $ C_q(\lambda) \overline{F}(z_{\rm Ly\alpha})$, just as they will contribute equally to the FFT computation described in the next section.

\subsection{Computation of  1D flux power spectrum}

To measure the 1D flux power spectrum $P_{1D}(k)$,  we express the quasar absorption spectrum in velocity units, where all pixels have the same size $\Delta v = 69 \;{\rm km/s}$.
We decompose each  spectrum $\delta_{\Delta v}$ into Fourier modes  and  estimate the variance as a function of wave number.  In practice, we compute the discrete Fourier transform of  the transmitted  flux  fraction  as described in~\citet{Croft1998}, using a fast Fourier Transform  (FFT) algorithm. In PYB13, we  developed in parallel a likelihood approach, in a  similar way to~\cite{McDonald2006}. As we  demonstrated that the two methods yield  compatible results, we will only pursue with a FFT approach in the present  analysis.  

The use of a FFT requires  the pixels to be equally spaced. The condition is satisfied with the  SDSS pipeline~\cite{Bolton2012} since the coadded spectra are computed with a constant pixel  width in velocity space. Throughout this paper we  therefore use velocity instead of observed wavelength. Similarly, the wave vector $k\equiv 2\pi/ v$ is measured in $\rm (km/s)^{-1}$. The  highest $k$-mode possible is determined by the Nyqvist-Shannon limit at $k_{\rm Nyqvist} = \pi/\Delta v$; with $\Delta v = 69 \;{\rm km/s}$,  the highest mode is $k_{\rm Nyqvist} = 0.045\;{\rm (km/s)^{-1}}$.  We limit the analysis, however,  to a maximal mode $k_{\rm max} = 0.02 \;{\rm (km/s)^{-1}}$, beyond which the spectrograph resolution cuts the power by over a factor of ten.  

In the absence of instrumental effects, the one-dimensional power spectrum can be simply written as the ensemble average over quasar spectra of $P^{raw}(k) \equiv   \left| \mathcal{F}(\delta_{\Delta v})  \right|^2$, where $\mathcal{F}(\delta_{\Delta v})$ is the Fourier transform of the transmitted flux  fraction $\delta_{\Delta v}$  in the quasar Ly$\alpha$ forest.
When taking into account the pixel noise $P^{noise}$, the impact of the spectral resolution of the spectrograph, the correlated  absorption $P^{{\rm Ly}\alpha-{\rm Si\,{III/II}}}$ of Ly$\alpha$ and either \ion{Si}{iii} or \ion{Si}{ii}, and the uncorrelated background $P^{metals}$ due to metal absorption such as  \ion{Si}{iv}  or  \ion{C}{iv},  the raw power spectrum is given by
\begin{equation}
P^{raw}(k)= \left( P^{{\rm Ly}\alpha}(k) +   P^{{\rm Ly}\alpha-{\rm Si\,{III/II}}}(k) +  P^{metals}(k)\right) \cdot W^2(k,R,\Delta v) + P^{noise}(k)
\label{eq:P1D_raw}
\end{equation}
where $W^2(k,R,\Delta v)$ is the window function corresponding to the spectral response of the spectrograph. The window function depends on the pixel width and on the spectrograph resolution $R$, such that
 \begin{equation}
 W(k,R,\Delta v)= \exp\left(- \frac{1}{2}(kR)^2\right) \times \frac{\sin(k\Delta v /2)}{(k\Delta v/2)}\,.
\label{eq:pk}
\end{equation}
The resolution is derived from  measurements obtained with spectral lamps, as described in~\cite{Smee2013}, and is provided by the eBOSS reduction pipeline~\cite{Bolton2012} for each coadded spectrum. Since the measurement of the 1D power spectrum on small scales is extremely sensitive to the  resolution, we adopt in this analysis the approach extensively described in~PYB13. This previous work provides a table of corrections of $R$ as a function of the position on the CCD in terms of fiber number and wavelength.  As in PYB13, we apply the correction to each pixel of each spectrum. 

The term $P^{{\rm Ly}\alpha-{\rm Si\,{III/II}}}(k)$ of  equation~\ref{eq:P1D_raw}, corresponding to the  correlated absorption by Ly$\alpha$ and \ion{Si}{iii} or \ion{Si}{ii} within the Ly$\alpha$ forest, can be estimated directly in the power spectrum. Since \ion{Si}{iii} absorbs at $\lambda_{\rm RF} = 1206.50\,$\AA, just 9\,\AA\  from Ly$\alpha$, its absorption appears in the power spectrum as oscillations with a frequency corresponding to  $\Delta v_{\rm Si\,{III}} \sim 2271 \, {\rm km/s}$.  Its contribution cannot be isolated from the Ly$\alpha$ absorption. Similarly,  absorption by \ion{Si}{ii} at $\lambda_{\rm RF} = 1190$ and 1193\,\AA\ creates an oscillatory pattern at a frequency corresponding to $\Delta v_{\rm Si\,{II}} \sim  5577\, {\rm km/s}$.  Both contributions are   included in the model of Ly$\alpha$ power spectrum that we fit to the data, in a similar way as was done in~PYB13. Note that metal lines aside from \ion{Si}{ii} and \ion{Si}{iii} with $1040<\lambda_{\rm RF}<1270$ are not detected and not accounted for in this analysis (cf.~\cite{Pieri2014} for metal lines that could have an impact). 

The other terms of equation~\ref{eq:P1D_raw}, the noise power spectrum $P^{noise}(k,z) $ and the metal power spectrum $P^{metals}(k)$, undergo specific updated treatments   compared to the  analysis of~PYB13, and are described in Sec.~\ref{sec:analysis}.

We compute the Fourier transform using the  efficient FFTW package from~\cite{Frigo2012}. The mean redshift of the Ly$\alpha$ absorbers in a forest determines the redshift bin to which the forest contributes. We rebin the final power spectrum onto an evenly spaced grid in $k$-space, assigning equal weight to the different Fourier modes that enter each bin. Using Eq.~\ref{eq:P1D_raw}, the final 1D power spectrum, $P_{1D}(k)$ is obtained by averaging the corrected power spectra of all  contributing forests, as expressed in the following estimator of  $P^{{\rm Ly}\alpha}(k)$:  
\begin{equation}
P_{1D}(k)  =  \left< \frac{ P^{raw}(k) - P^{noise}(k) }{W^2(k,R,\Delta v)} \right> \,  -  P^{{\rm SB_1}}(k),
\label{eq:P1D_FFT}
\end{equation}
where  $\langle\rangle$ denotes the ensemble average over forest spectra and where  $P^{{\rm SB_1}}(k)$ is the power spectrum measured in the first side band corresponding to the wavelength range  $ 1270 < \lambda_{\rm RF}< 1380 \,\AA$ as shown in Fig.~\ref{fig:Spectra_eBOSS}.  In  Eq.~\ref{eq:P1D_FFT}, $P^{{\rm SB_1}}$ includes the power from uncorrelated metals  $P^{metals}$ but also a contaminating contribution coming from the spectroscopic pipeline. Details are given in Sec.~\ref{sec:SB}. The final result is presented in 35 evenly spaced $k$-modes with $\Delta k=5.4\times 10^{-2}\,\rm (km/s)^{-1}$,  in 13 evenly spaced redshift bins from $z_{\rm Ly\alpha}=2.2$ to 4.6.

\section{Data  analysis}
\label{sec:analysis}
\subsection{Sample selection}
\label{sec:selection}

The DR14Q quasar catalog contains 525,982 quasars. We are solely interested in the 209,407 quasars  observed by BOSS or eBOSS at  $z_{\rm qso}>2.1$ for which the \lya\ region is accessible. When restricting to visually inspected spectra (those with ${\rm MJD}<56870$), the sample reduces to 180,413 objects. Finally, we discarded quasars with BAL features, as flagged by a non-zero value of BI\_CIV, bringing the initial sample to 167,988 quasars. In case of multiple observations of a quasar, we only use the best one. 

The analysis of PYB13 was  reaching a similar level of statistical and systematic uncertainties on some modes or redshifts, despite being based upon an initial sample of only about 60,000 $z_{\rm qso}>2.1$ quasars, from which about $14,000$ were selected. The significant increase in statistics of the present sample compared to PYB13 makes it possible, if not mandatory, to tighten the selection criteria in order to reduce the impact  of the systematic uncertainties on the measured power spectrum. The following criteria are applied on each forest, and no longer on the selection of the quasar spectra. 

\begin{figure}[htbp]
\begin{center}
\epsfig{figure= 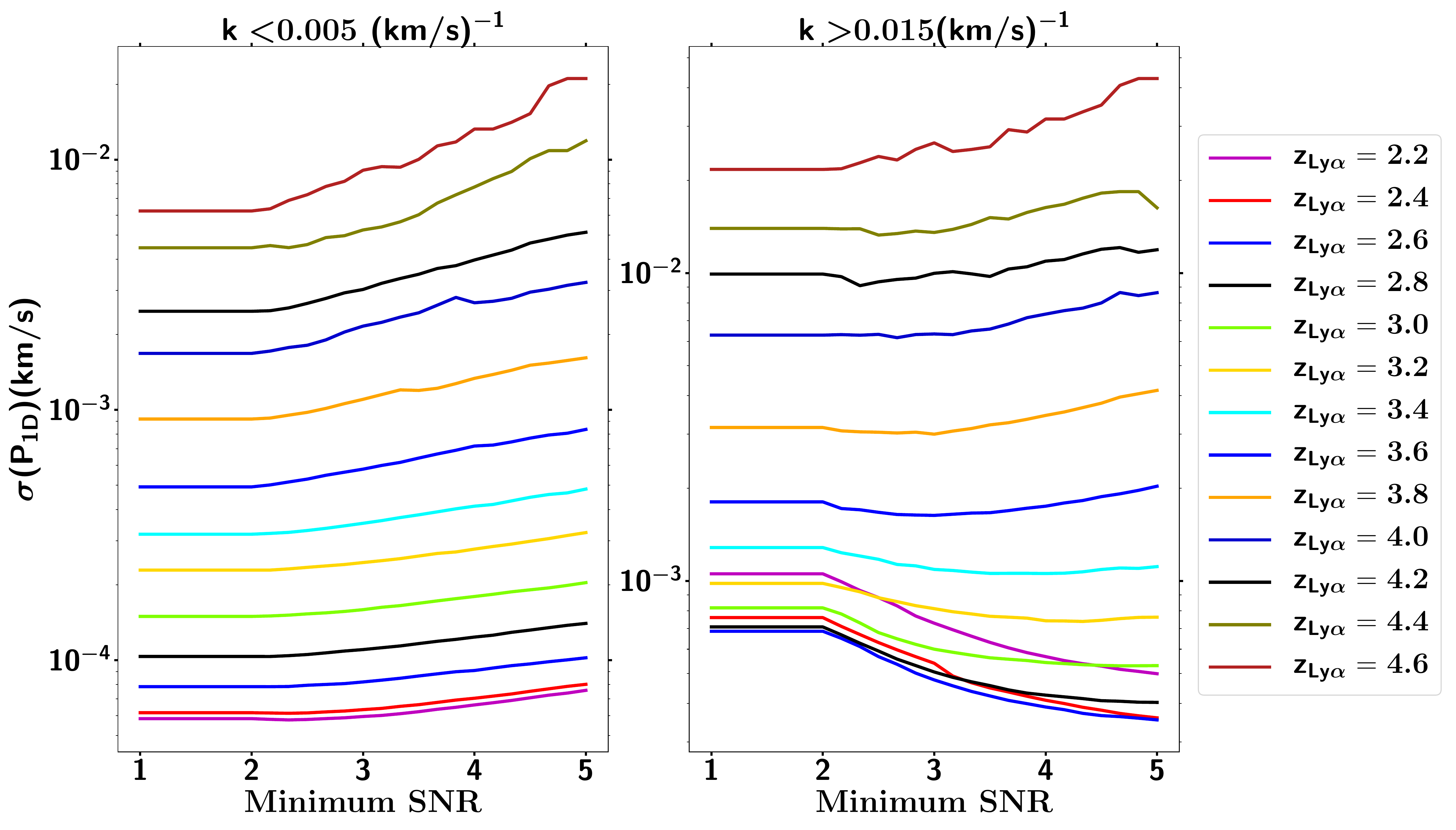,width = \textwidth}\\
\caption{\it Uncertainty on the mean value of $P_{1D}$ for the 13 redshift bins, focusing  on the average over either large scales ($k<0.005$, left panel) or small scales ($k>0.015$, right panel). } 
\label{fig:SNR}
\end{center}
\end{figure}

To improve the quality of the low-redshift forests, we increased the CCD short-wavelength cut, below which the CCD becomes considerably noisier, from 3650 to 3750\,\AA.  We discard forests where the mean spectral resolution is larger than $85\,{\rm km/s}$. We also optimize the threshold on the mean signal-to-noise ratio per pixel (SNR) below which we reject a forest, where the SNR is defined as the ratio of the pixel flux to the pixel noise, and the average is computed after pixel masking. 
This optimization is done by computing, for each redshift bin, the uncertainty on the mean value of $P_{1D}$, as a function of the threshold on the mean SNR per pixel. This uncertainty depends upon the range of modes considered, as illustrated in Fig.~\ref{fig:SNR} where we test the $P_{\rm 1D}$ uncertainty in samples of varying minimum SNR. 
On large scales where  signal dominates, the uncertainty decreases with more statistics from a more inclusive sample, and hence a lower threshold, whereas on 
small scales where  noise tends to dominate, the higher the threshold the lower the uncertainty. We therefore set the SNR threshold so as to minimize the uncertainty around a central mode, at $k\sim 0.01\,\rm(km/s)^{-1}$. We checked that our final power spectrum measurement does not contain significant variations with a $\pm 0.3$ shift in the chosen SNR threshold. For  $z_{\rm Ly\alpha}\ge3.4$, the uncertainty  no longer evolves significantly with SNR, and the threshold is fairly independent of redshift. As a result, we apply the  thresholds given in table~\ref{tab:SNR}  for the forest selection. For comparison, it was set to 2.0 for all redshifts in PYB13.

\begin{table}[htbp]
\caption{Threshold on the minimum mean signal-to-noise ratio per pixel in a forest}
\begin{center}
\begin{tabular}{lcccccccc}
\hline
\hline
$\left< z_{\rm Ly\alpha} \right>$ & 2.2 & 2.4& 2.6& 2.8& 3.0& 3.2 & 3.4 & 3.6 to 4.6\\
SNR threshold & 4.1 & 3.9 & 3.6 & 3.2 & 2.9 & 2.6 & 2.2 & 2.0 \\
\hline
\end{tabular}
\end{center}
\label{tab:SNR}
\end{table}

Sky lines  affect the data quality by  increasing the pixel noise. We mask major sky lines (e.g., lines at 5577\,\AA, 5890\,\AA, 6300\,\AA, 6364\,\AA, 6864\,\AA)   in each forest.
 We locate the position of DLAs in our forest using catalog N12 introduced in Sec.~\ref{sec:vac}. We mask the DLAs following the same procedure as in~\cite{Bautista2017}: all pixels where the DLA absorption  is higher than 20\% are masked, and the absorption in the wings is corrected  using a Voigt profile.

We discard forests shorter than 75 pixels, where the reduced length can be due  to the CCD UV cut  or to the presence of a strong DLA at the forest boundary. We also discard forests with more than 40 masked pixels, whether  from DLA absorption,  sky line masking, or flags from the SDSS pipeline (the latter are indicated by a null variance). Since the use of a FFT to compute the power spectrum requires equally-spaced pixels, we reintroduce all masked pixels in the forest before performing the Fourier transform and set their flux to zero. This procedure introduces a  $k$-dependent bias in the resulting power spectrum, which we quantify and correct  as discussed in Sec.\ref{sec:mocks}.

\begin{figure}[htbp]
\begin{center}
\epsfig{figure= 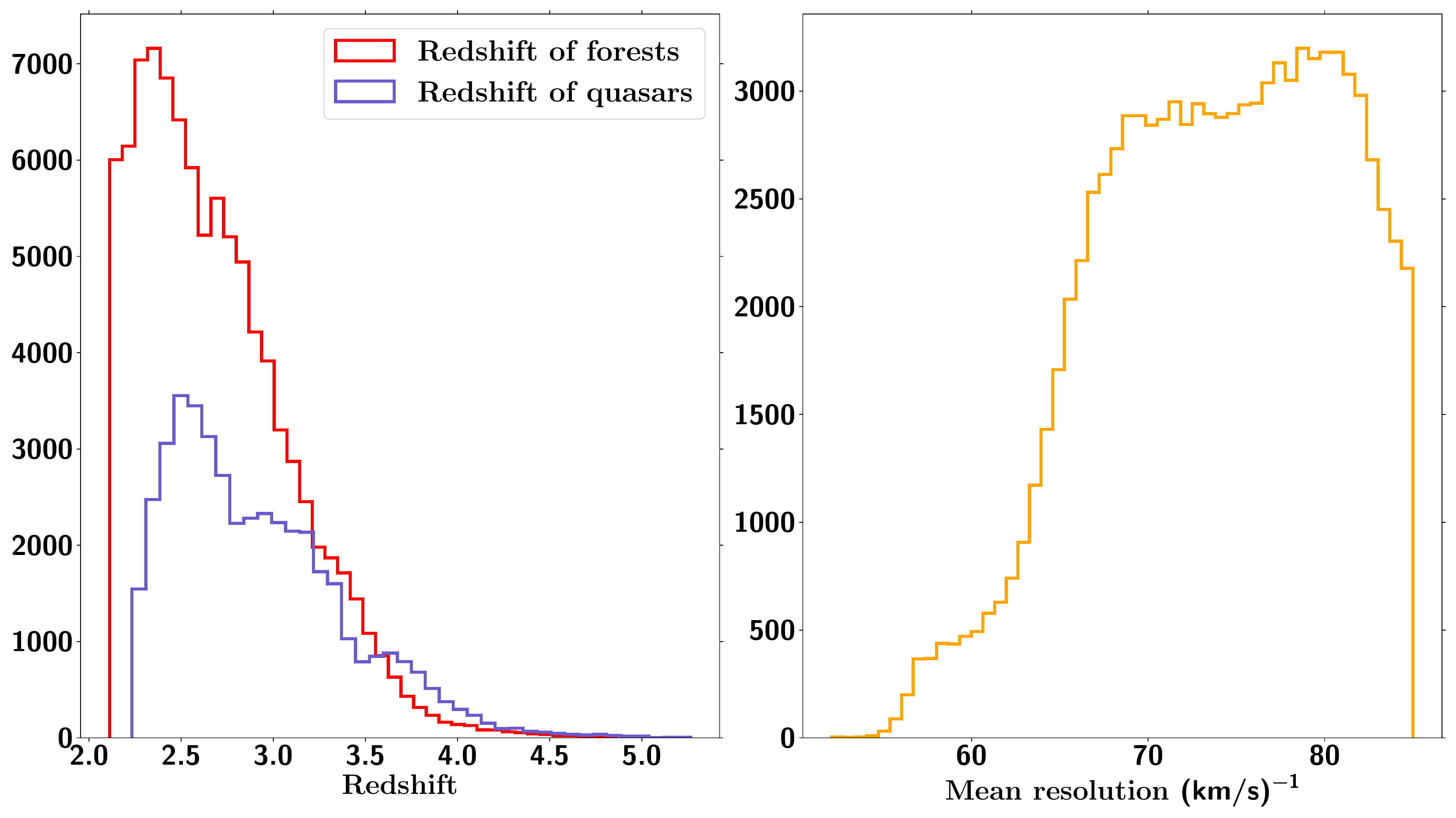,width = \textwidth}\\
\caption{\it Distributions of mean forest  and mean quasar redshift on the left, and mean resolution per pixel on the right for the selected forest sample.} 
\label{fig:nz_reso}
\end{center}
\end{figure}

This tight  selection procedure yields a sample a 43,751 high-quality quasar spectra,  from which we use 94,558~forests\footnote{Recall that {\em forests} are  sub-regions defined in Sec.~\ref{sec:transmittedflux} of  the commonly-called Ly$\alpha$ forest of quasar spectra.}  for the analysis. The yields per redshift bin are summarized in table~\ref{tab:yields}.  The UV cut, the resolution cut, and the stringent SNR cut at low redshift contribute to often discarding the first (or even first two) forests of a given quasar spectrum, since the blue end of the spectrograph suffers from all these drawbacks (large noise and poor resolution). These cuts explain why the number of forests is not  simply three times the number of quasars. Fig.~\ref{fig:nz_reso} presents the resulting redshift and resolution distribution of our selected forests. 

\begin{table}[htbp]
\caption{Summary per redshift bin of the number of forests, the mean redshift, the mean SNR per pixel, and the mean resolution.}
{\footnotesize
\begin{center}
\begin{tabular}{cccccccccccccc}
\hline
\hline
z bin & 2.2 & 2.4& 2.6& 2.8& 3.0& 3.2 & 3.4 & 3.6 & 3.8& 4.0& 4.2& 4.4& 4.6\\
\# & 17,144& 20,089 & 16,541 & 14,762 & 10,364 & 6,767 & 4,763 & 2,356 & 933 & 421 & 229 & 126 & 63 \\
$\left< z_{\rm Ly\alpha} \right>$ & 2.207 & 2.396& 2.595& 2.795& 2.991& 3.190 & 3.393 & 3.587 & 3.786& 3.994& 4.194& 4.387& 4.578\\
$\left< {\rm SNR} \right>$ & 7.3 & 7.6 & 7.3 & 6.8 & 6.5 & 6.0 & 5.2 & 4.6 & 4.3 & 3.9 & 4.0 & 3.8 & 3.5 \\
$\left< {\rm R} \right>$ & 81.2 & 77.8 & 73.9 & 71.0 & 68.9 & 67.3 & 66.1 & 65.2& 65.9 & 70.6& 72.9 & 71.2 & 69.3 \\

\hline
\end{tabular}
\end{center}
}
\label{tab:yields}
\end{table}

\subsection{Estimator of noise power spectrum}
\label{sec:noise}
A quasar spectrum is typically measured with five to eight individual exposures. We use these  $N_{\rm exp}$ individual exposures to compute the noise power spectrum $P^{noise}$  in a two-step process. 

We first construct a spectrum that contains the same noise as the data coadded spectrum, but devoid of any  power from an astrophysics or cosmology signal. To this end, we
compute the semi-difference $\Delta\phi$ between two customized coadded spectra  obtained from weighted averages of the even-number exposures, for the first spectrum, and of the odd-number exposures, for the second one, with weights taken as the pixel inverse variances.  In this first step, we assign zero  weight to pixels flagged by the  COMBINEREJ bit of the SDSS pipeline, i.e., pixels rejected when computing the coadded spectrum from the  individual exposures. 
Even assuming all exposures to have the same variance, the variance measured in $\Delta\phi$ is the same as the variance in the data coadded spectrum only if $N_{\rm exp}$ is even. In contrast, if $N_{\rm exp}$ is odd, we can write $N_{\rm eff}=2N_1+2$, and the variance of $\Delta\phi$ is reduced by $[1/N_1 + 1/(N_1+1)]/4$ with respect to the variance in each exposure, instead of by the  $1/N_{\rm exp}$ factor that is expected for the data coadded spectrum.   
To account for a possible difference in the number of exposures of the two customized spectra, we thus renormalize $\Delta\phi$ by the factor $\sqrt{\alpha}$, where $\alpha = 4\times E[N_{\rm exp}/2)]\times E[(N_{\rm exp}+1)/2)]/N_{\rm exp}^2$, with $E$ denoting the truncated integral part. This approach gives $\alpha=1$ if $N_{\rm exp}$ is even, and, for instance, $\alpha= 0.98$ for a quasar with five exposures. The quantity $P_{\rm diff}^{noise}  (k)= | \mathcal{ F} (\alpha \Delta\phi) |^2$, where $\mathcal{ F} (\alpha \Delta\phi)$ is the Fourier transform of the normalized difference spectrum, is expected to be an accurate estimator of the noise power in the coadded spectrum. $P_{\rm diff}^{noise}  (k)$ is found to be scale-independent to an accuracy sufficient for our purpose, as expected for a white noise. We thus define $P_{\rm diff}^{noise}$ as the average of $P_{\rm diff}^{noise}(k)$ over all $k$-modes covered by our analysis.

\begin{figure}[htbp]
\begin{center}
\epsfig{figure= 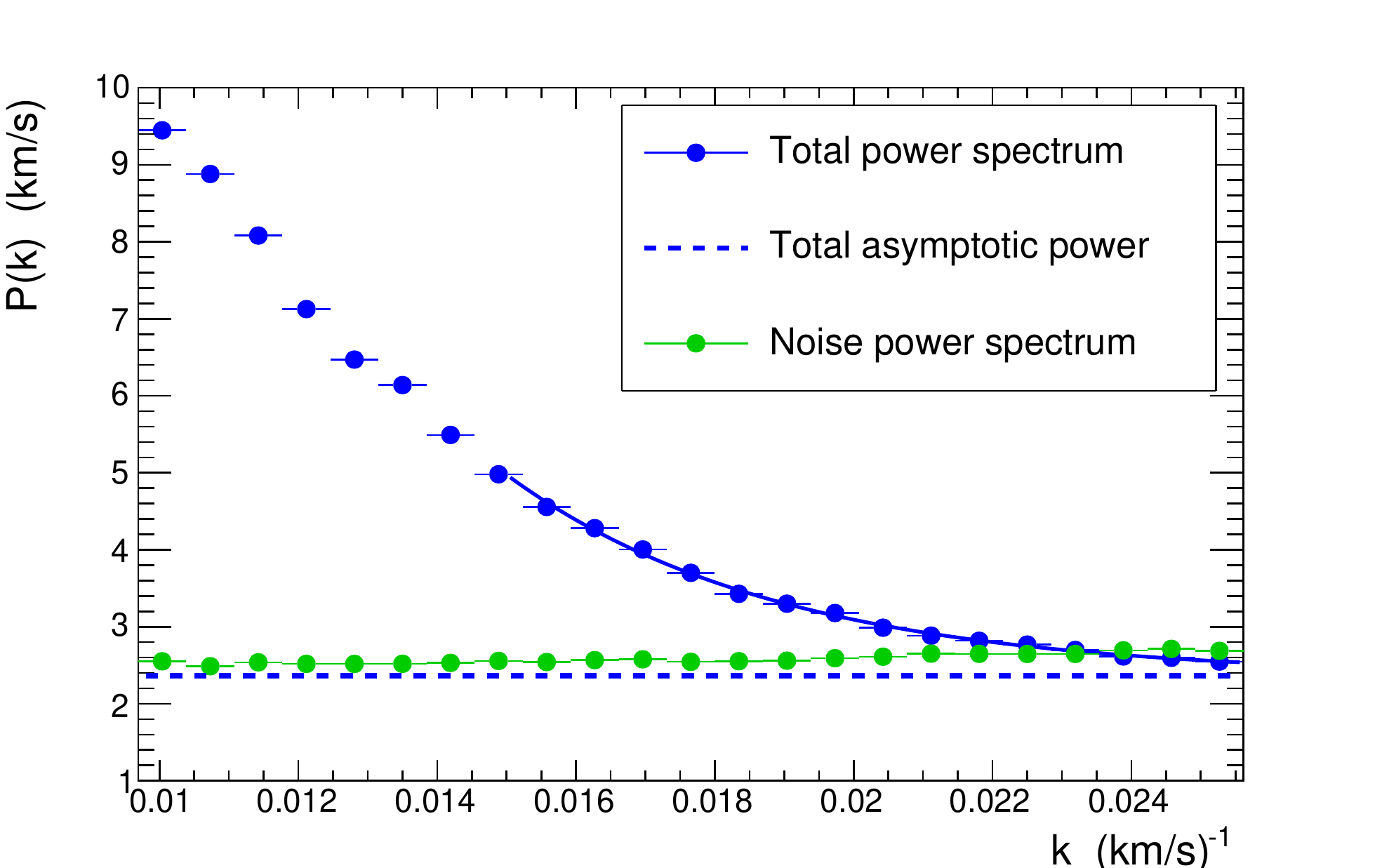,width = 14cm}\\
\caption{\it  Power spectra  of the noise (green circles),   raw data (blue circles), and the asymptote of the total power spectrum (dashed blue line),  for $z<2.5$. A correction is applied to the noise level when the latter exceeds the raw power spectrum. } 
\label{fig:noise}
\end{center}
\end{figure}

Because the spectrograph window function  $W^2$ quickly drops to zero beyond the scales used in the analysis, we can further refine the estimate of the noise power spectrum by checking that on very small scales, $P_{\rm diff}^{noise}$ is indeed a lower asymptote of the raw power spectrum $P^{raw}$, as expected from Eq.~\ref{eq:P1D_raw}. Fig.~\ref{fig:noise} shows that this is not the case at small redshifts, due to subtle differences between the pipeline coaddition and the above semi-difference computation. In a second step, for each of the 13 redshift bins, we therefore fit $P^{raw}$ on $k$-modes above $k_{\rm max} = 0.02\,{\rm (km/s)^{-1}}$ by an exponential decrease plus a constant $P^{raw}_{\rm lim}$ (shown as the blue dashed line in Fig.~\ref{fig:noise}).
We compute the ratio $\beta$ between $P_{\rm diff}^{noise}$ and $P^{raw}_{\rm lim}$. We define $P^{noise}=P_{\rm diff}^{noise}$ for redshift bins where $\beta<1$, and  $P^{noise}=P_{\rm diff}^{noise}/\beta$ otherwise. We find $\beta$ of order 0.95 for the first three redshift bins, and we set it to 1.0 for higher redshifts.

\subsection{Side-bands}
\label{sec:SB}

The metal power spectrum $P^{metals}(k)$ of equation~\ref{eq:P1D_raw}, corresponding to uncorrelated background due to metal absorption in the Ly$\alpha$ forest, is independent of  Ly$\alpha$ absorption and  cannot be estimated directly from the power spectrum measured in the Ly$\alpha$ forest. PYB13  addressed this issue by estimating the power spectrum  in side bands located at longer wavelengths than the  Ly$\alpha$ forest region, and the power spectrum was subtracted  from the Ly$\alpha$ power spectrum measured in the same gas redshift range. This method presents the advantage of allowing us  to simultaneously account for most residual effects affecting our determination of the 1D power spectrum. We   apply the same technique in this paper. 

We define two side bands, ${\rm SB_1}$ and ${\rm SB_2}$, corresponding, respectively,  to the wavelength range  $ 1270 < \lambda_{\rm RF}< 1380 \,\AA$  and $1410 < \lambda_{\rm RF}< 1520 \,\AA$  as shown in Fig.~\ref{fig:Spectra_eBOSS}. The  power spectrum measured in the first side band,  ${\rm SB_1}$, contains the complete contribution from all metals with  $\lambda_{\rm RF}>1380\,$\AA, including in particular absorption from \ion{Si}{iv} and \ion{C}{iv}. The second side band, ${\rm SB_2}$ also includes  \ion{C}{iv},  but not the \ion{Si}{iv} absorption.  We thus use  ${\rm SB_1}$  to subtract the metal contribution in the power spectrum,  and   ${\rm SB_2}$  as an important consistency check. 

Our method is purely statistical: for a given redshift bin, we use different quasars to compute  the Ly$\alpha$ forest and the metal power spectra. For instance, for the first redshift bin,  $2.1<z_{\rm Ly\alpha}<2.3$, we measure the power spectrum in ${\rm SB_1}$,  corresponding to  $ 3750 < \lambda< 4011\,$\AA, i.e., using quasars with a redshift $z_{\rm qso}\sim1.9$. Quasars in a given redshift window have their two side-bands corresponding to fixed observed wavelength windows, which in turn match a specific redshift window of Ly$\alpha$ forest.

\begin{figure}[htbp]
\begin{center}
\epsfig{figure= 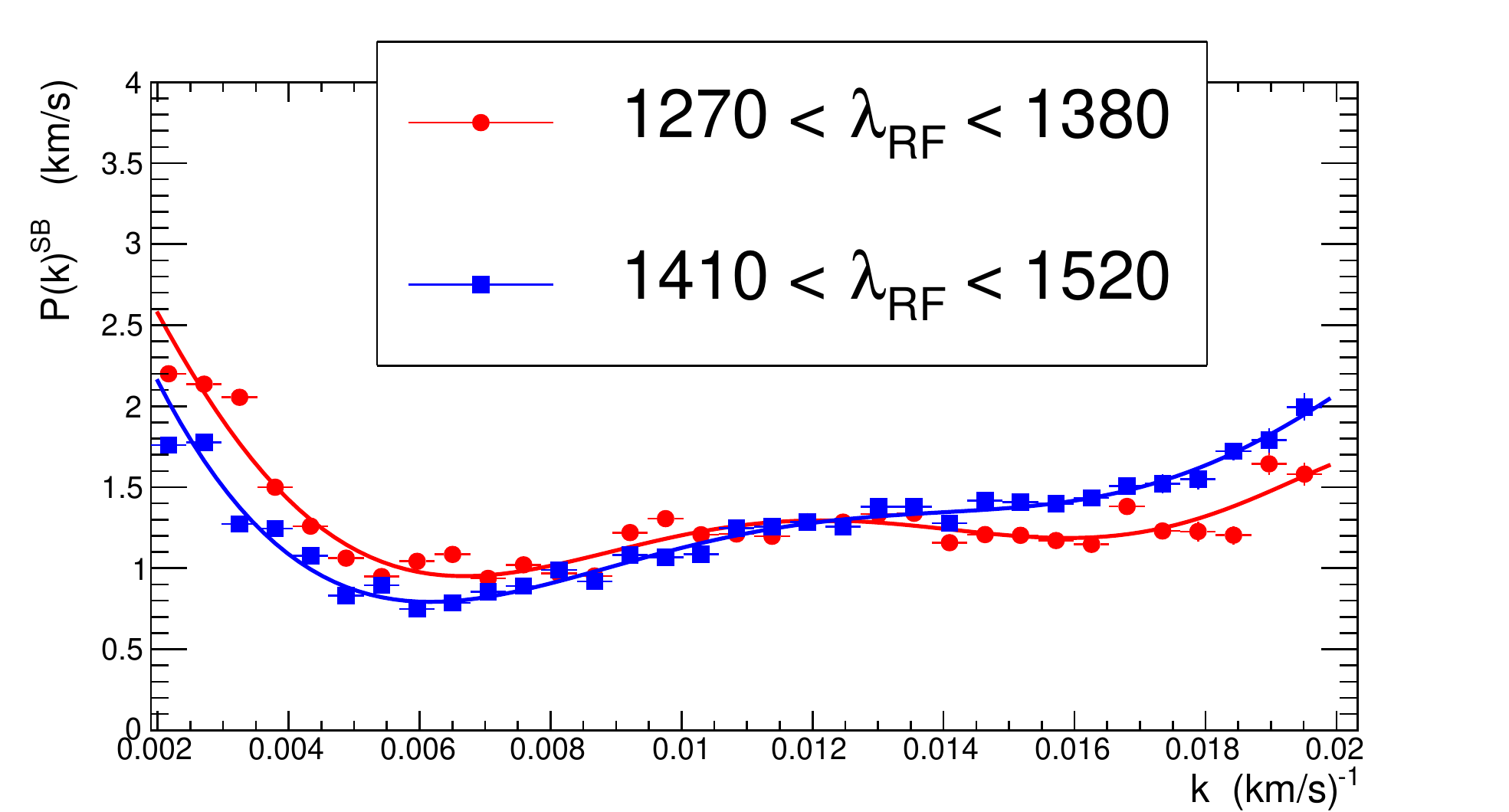, width = 14cm}
\caption{\it Power spectrum $P^{\rm SB}(k)$ computed for side band regions on the red side of the Ly$\alpha$ forest. The red dots  and the blue squares represent the two side bands defined  by $ 1270 < \lambda_{\rm RF}< 1380\,${\rm \AA} and $ 1410 < \lambda_{\rm RF}< 1520\,${\rm \AA}, respectively. Each power spectrum is fitted with a  sixth-degree polynomial. } 
\label{fig:TwoSB}
\end{center}
\end{figure}
\begin{figure}[htbp]
\begin{center}
\epsfig{figure= 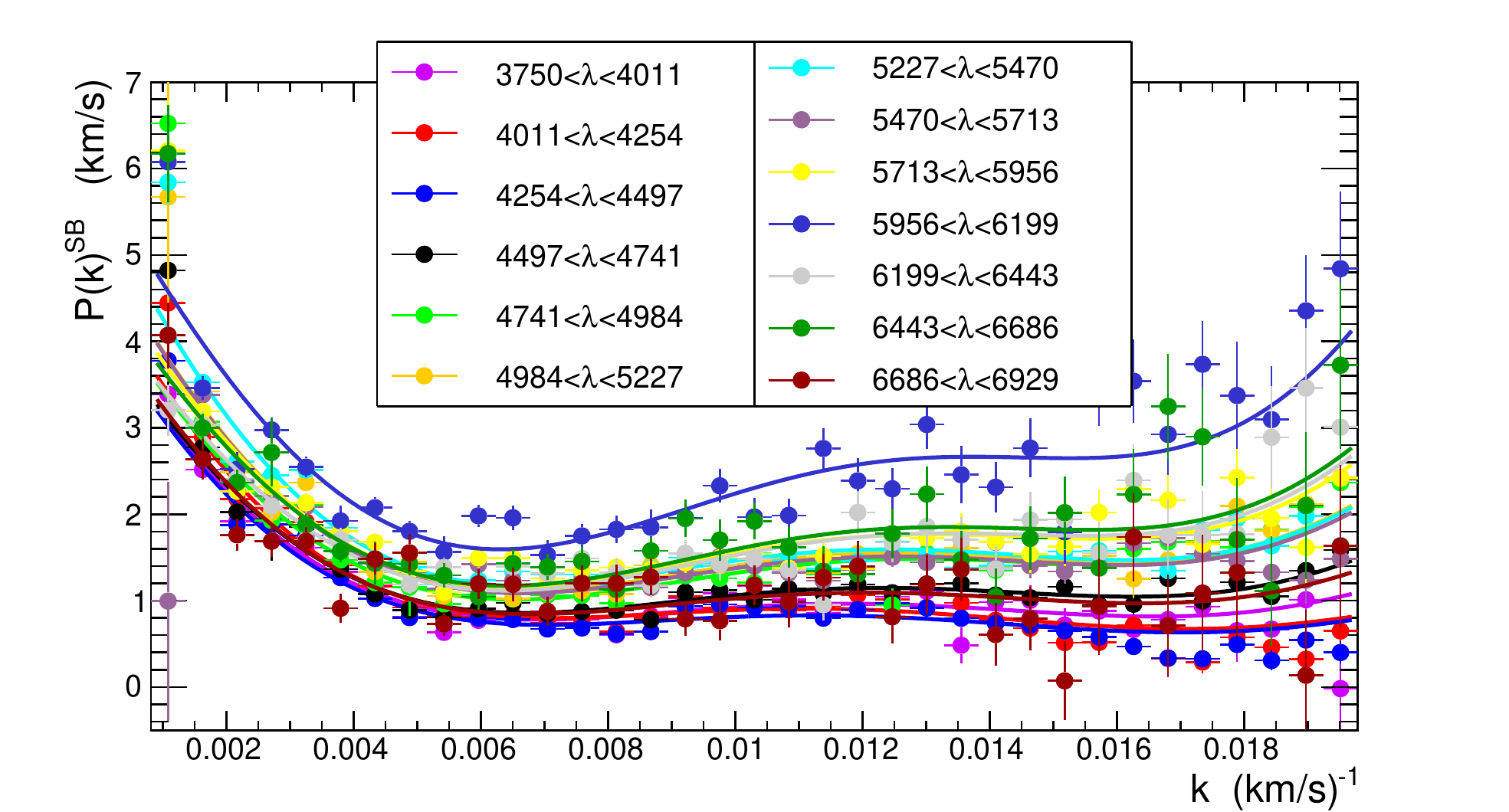, width = 14cm}
\caption{\it Power spectrum $P^{\rm SB_1}(k)$ computed for the first side band region $ 1270 < \lambda_{\rm RF}< 1380\,{\rm \AA}$ redward of the Ly$\alpha$ forest  for the 13 different $\lambda$ windows. Each $\lambda$ region corresponds to one redshift bin.  Each power spectrum is fitted by the product of the sixth-degree polynomial obtained in Fig.~\ref{fig:TwoSB} and a first-degree polynomial in which the two parameters are free.} 
\label{fig:SB1_zbins}
\end{center}
\end{figure}

The power spectra $P^{\rm SB}(k)$ shown in Fig.~\ref{fig:TwoSB}  are obtained, respectively, for ${\rm SB_1}$ and ${\rm SB_2}$ with  about 115,000 and   140,000 quasars passing similar quality cuts as the quasars selected for the Ly$\alpha$ forest analysis.  The shapes of $P^{\rm SB}(k)$ are similar for the two side bands. As expected, for ${\rm SB_2}$, which excludes \ion{Si}{iv}, the magnitude of  $P^{\rm SB}(k)$  is smaller for $k<0.01 {\rm (km/s)^{-1}}$.  On smaller scales, the shape is dominated by  residuals effects and systematics that exceed the strength of the metal absorption.  
On small scales, SB2 exhibits more power than SB1 because the measurement includes more quasar spectra affected by the excess of power in the $\lambda\sim 6000\,\AA$ region, as explained below.
We fit the distribution $P^{\rm SB}(k)$ with a sixth-degree polynomial. We will use this fitted function as a template to parametrize the $P^{\rm SB}(k)$ measured for each wavelength window (see  Fig.~\ref{fig:TwoSB}). 

As the shape and the magnitude of the power spectrum vary from one  wavelength window to another, we  parameterize $P^{\rm SB}(k)$ as the product of  the fixed shape of Fig.~\ref{fig:TwoSB} by a  first-degree polynomial with two free parameters that  differ for each wavelength window. This model adequately fits the measured power in all the wavelength windows (see Fig.~\ref{fig:SB1_zbins}). From these parametric functions, we extract the value of the power spectrum $P^{\rm SB_1}(k)$  for each $k$ and for each Ly$\alpha$ redshift window.

As mentioned above, Fig.~\ref{fig:SB1_zbins} suggests that  $P^{\rm SB_1}(k)$ does not decrease to zero at small scales, unlike what would have been expected  due to thermal broadening, just as for the Ly$\alpha$ power spectrum. This effect is more pronounced for the three redshift bins that contain  the overlap between the two arms of the spectrograph, $ 5800 < \lambda< 6350 \,\AA$, indicating that the excess is likely due to imperfections in the co-addition between the two parts of the spectrum.  This behavior also occurs for side bands as well as for  \lya. Therefore,  subtracting the power spectrum computed in the side bands not only  removes the contribution due to metal absorption but also corrects for residual effects of the pipeline. In the most dramatic case ($\lambda\sim 6000 \,\AA$), the residual effect measured at high $k$ corresponds only to $\sim 10\%$ of the power spectrum measured in the  Ly$\alpha$ forest. This correction leads to a 3\% systematic uncertainty (cf. Sec.~\ref{sec:systs}), which is small compared to the statistical error bar.


\section{Synthetic data and bias corrections}
\label{sec:mocks}
In this section, we investigate the biases introduced at each step of the data analysis, and estimate their impact using mock spectra. They arise from the estimated value of $C_q(\lambda) \overline{F}(z_{\rm Ly\alpha})$ and from the masking of pixels affected by sky emission lines or absorption by DLAs.

\subsection{Mocks}
\label{sec:mock}
To test the analysis procedure and investigate systematic errors, we generated mock spectra that reproduce the essential physical and instrumental characteristics of the eBOSS spectra. The mocks are produced following the procedure described below. First, a redshift and a $g$-magnitude are chosen at random from  distributions tuned to data. Second, an unabsorbed flux spectrum is drawn for each quasar from a random selection of PCA amplitudes following the procedure of~\cite{paris11}, and the flux normalized to  the selected $g$ magnitude. Third, the Ly$\alpha$ forest absorption is generated following a procedure adapted from~\cite{font-ribera12}, who provide an algorithm for generating any spectrum of the transmitted flux fraction $F(\lambda)$ from a Gaussian random field $g(\lambda)$. Specifically, they present a recipe for choosing the parameters $a$ and $b$ and the power spectrum $P_g(k)$ such that the transformation $F(\lambda)=\exp[-a\exp(bg(\lambda))]$ yields the desired power spectrum and mean value of $F(\lambda)$. In practice we generate a suite of transmitted-flux-fraction spectra for thirteen redshifts that reproduce the observed power. For each wavelength pixel,  $F(\lambda)$ is  obtained by interpolation between redshifts according to the actual Ly$\alpha$ absorption redshift of the pixel. The unabsorbed flux is  multiplied by $F(\lambda)$ and convolved with the spectrograph resolution. The spectra are  generated with a pixel width that is three times small than a SDSS pixel, and about three times smaller than the SDSS spectral resolution. We checked that this size was  small enough to properly take into account the spectral resolution. Finally,  noise is added according to eBOSS throughput and sky noise measurements as was done in~\cite{legoff11}, and the spectrum is rebinned  to the SDSS format. 
We generated two sets of mock spectra. In the first one, the quasars distribution is the same as the one resulting from our sample selection described in~\ref{sec:selection}. The second set contains twenty times the quasar distribution of redshift z $>$ 3.7 to improve statistical uncertainty in the high-redshift bins.

\subsection{Continuum estimation effect}\label{sec:cont}
\label{sec:bcont}
As a  starting point, we checked that computing $\delta(\lambda)$ of Eq.~\ref{eq:delta} with the generated values of the quasar continuum, $C_q(\lambda)$, and of the mean transmitted flux, $\overline{F}(z_{\rm Ly\alpha})$, allows an accurate reconstruction of the input power in the absence of noise and resolution effects. This step validates the implementation of the code that computes the power spectrum of a $\delta$ field. 

\begin{figure}[htbp]
\begin{center}
\epsfig{figure= 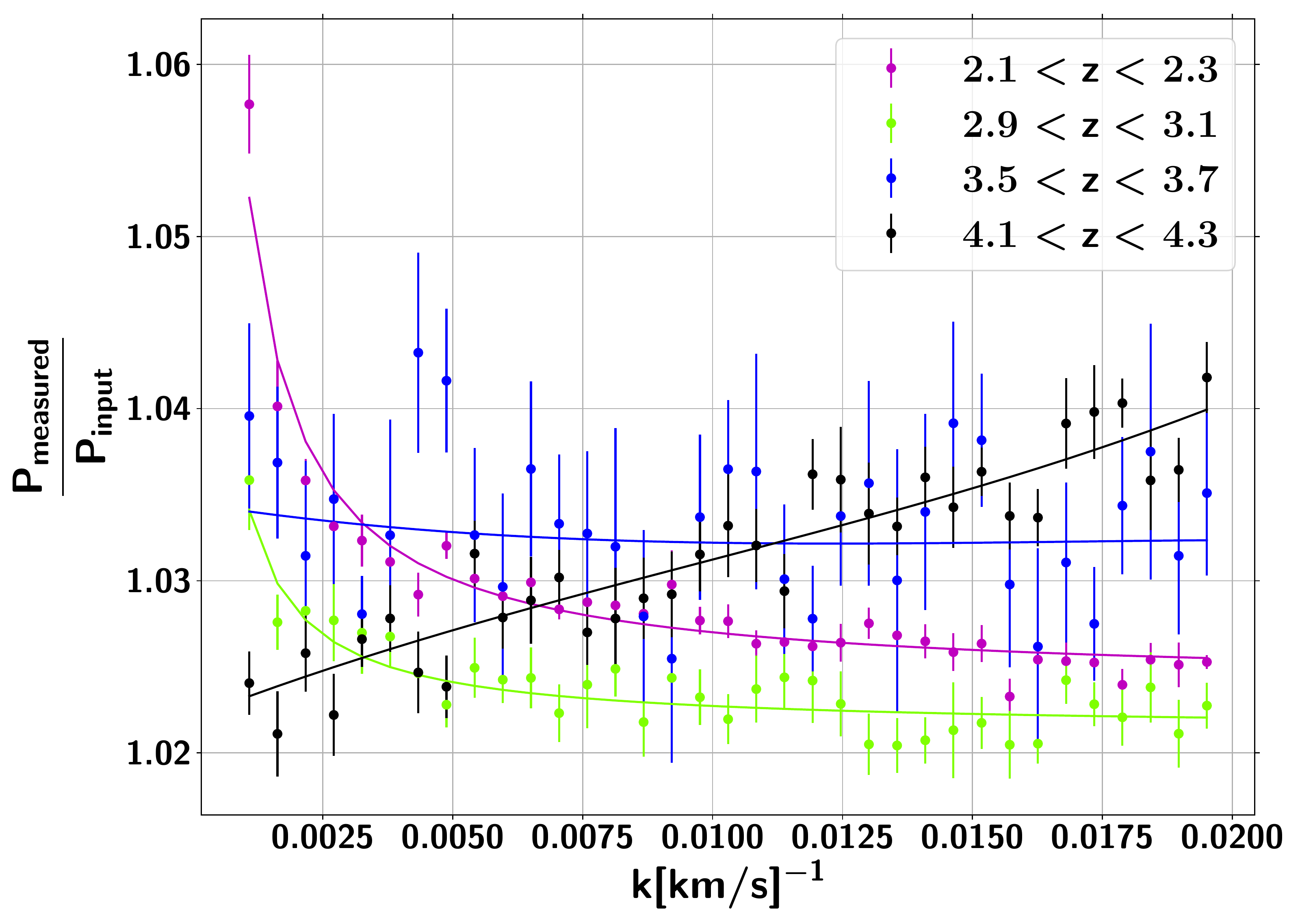, width = \textwidth}
\caption{\it Ratio of measured power spectrum $P_{measured}$ to the flux  spectrum that was generated in the mock spectra $P_{input}$, illustrated for four redshift bins. The correction is fitted with a functional $a/k + b$ dependence for $2.1<z_{\rm Ly\alpha}<3.5$ and with a $3^{rd}$-degree polynomial for $3.5<z_{\rm Ly\alpha}<4.7$.}
\label{fig:ContBias}
\end{center}
\end{figure}

 We then use the value of $C_q(\lambda)\overline{F}(z_{\rm Ly\alpha})$ that we estimate as explained is Sec.~\ref{sec:transmittedflux}.
We define the bias induced by  the continuum  estimation as the ratio of the measured flux power spectrum, $P_{measured}$, to the flux power spectrum that was generated in the mock spectra, $P_{input}$,
\begin{equation}
b(k,z) = \frac{P_{measured}(k,z)}{P_{input}(k,z)}.
\label{eq:cont_bias}
\end{equation}
The mocks were tuned so that $P_{input}(k,z_{\rm Ly\alpha})$ and $\overline{F}(z_{\rm Ly\alpha})$ reproduce the data values. Figure~\ref{fig:ContBias} illustrates the measured bias.  

The use of Eq.~\ref{eq:continuum} for the continuum instead of the real quasar continuum introduces correlated noise in the estimate of $\delta(\lambda)$, which results in a bias larger than unity. The shape of this bias as a function of $k$ depends on the relative amplitudes of the LSS power spectrum and that of this correlated noise, and therefore evolves with  redshift.

We also tested a quasar-dependent term of the form  $a_q+b_q(\lambda_{RF}-\overline{\lambda}_{RF})$ where $\overline{\lambda}_{RF}$ is the mean over the forest. We validate the method principle by verifying that  whatever the form used for the quasar-dependent term, the power spectra of the data after correction by the relevant correction functions (via Eq.~\ref{eq:cont_bias}) yield consistent results, with less than 1\% difference between the two  options. The latter form  yields a larger bias than the simpler multiplicative constant, so we do not consider it further.

\subsection{Pixel masking effect}
\label{sec:bmasking}
We mask pixels  affected by strong absorption caused by DLA or by emission from sky  lines,  by setting their flux to zero. Otherwise,  sky lines would impact the data quality by increasing significantly the pixel noise, and  DLAs, responsible for saturated absorptions on the scale of several pixels, would generate additional  correlations. We evaluate the impact of  pixel masking by applying the same  procedure on mock spectra that  include neither sky lines nor DLAs, and we measure the ratio:
\begin{equation}
b(k,z) = \frac{P_{m}(k,z)}{P_{u}(k,z)},
\label{eq:mask_bias}
\end{equation}
with $P_{m}$ the masked power spectrum  and $P_{u}$ the unmasked power spectrum. In order to only evaluate the masking impact, both power spectra are computed using the generated values for the quasar continuum $C_q(\lambda)$ and for the mean transmitted flux $\overline{F}(z_{\rm Ly\alpha})$.

\begin{figure}[htbp]
\begin{center}
\epsfig{figure= 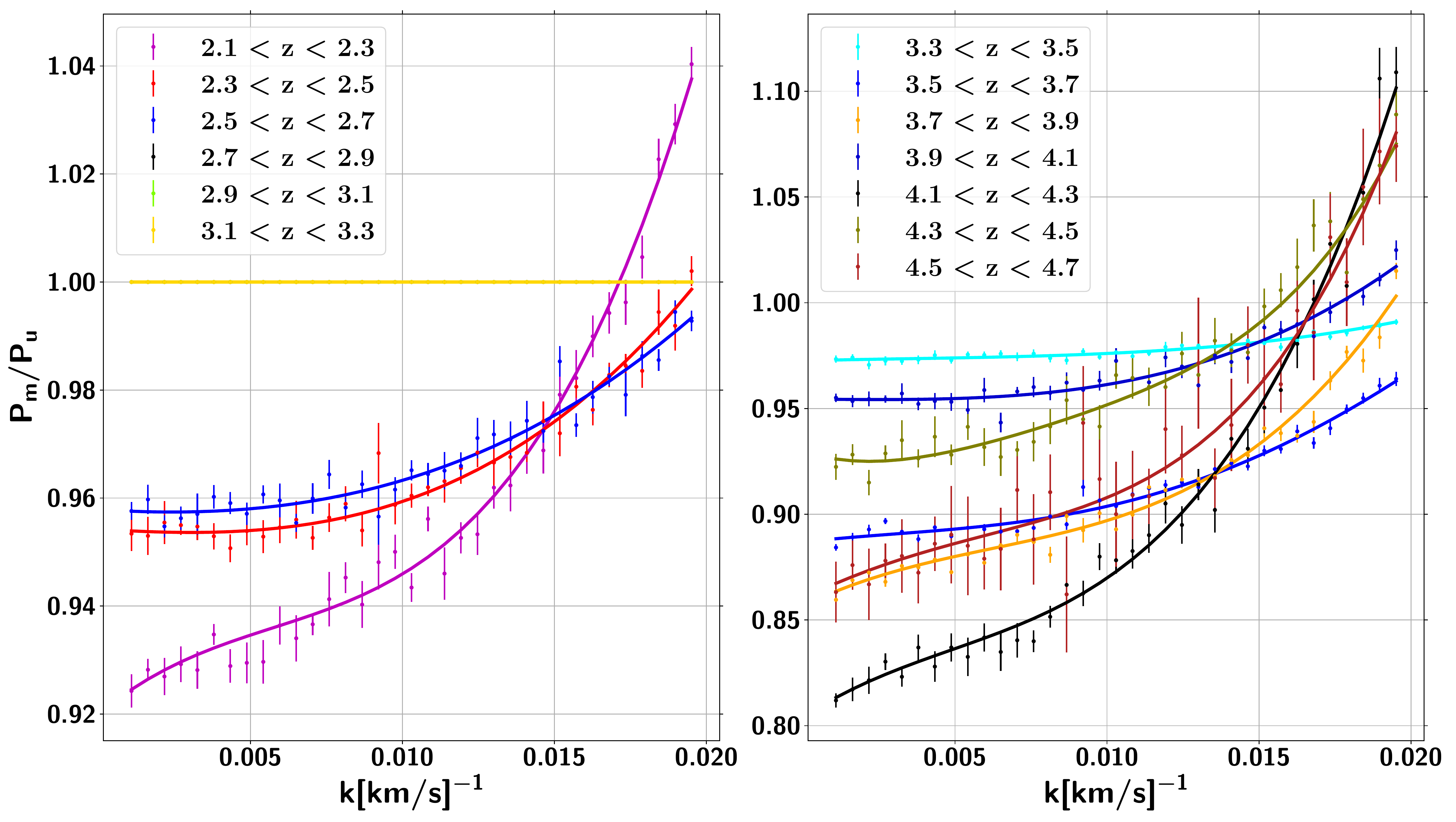,width = .97\textwidth}
\caption{\it Ratio of  the masked power spectrum $P_{m}$ to the unmasked power spectrum $P_{u}$, due to sky emission line masking, for all thirteen redshift bins from $z_{\rm Ly\alpha}=2.2$ to 4.6.  No strong sky line enters the forest in 2.7 $< z_{\rm Ly\alpha} <$ 3.3, resulting in no correction in this redshift range. The bias in 4.3 $< z_{\rm Ly\alpha} <$ 4.7 and 3.5 $<z_{\rm Ly\alpha} <$ 3.7 is modeled by a $4^{th}$-degree polynomial, and in the other redshift bins by a $3^{rd}$-degree polynomial. } 
\label{fig:SklBias}
\end{center}

\begin{center}
\epsfig{figure= 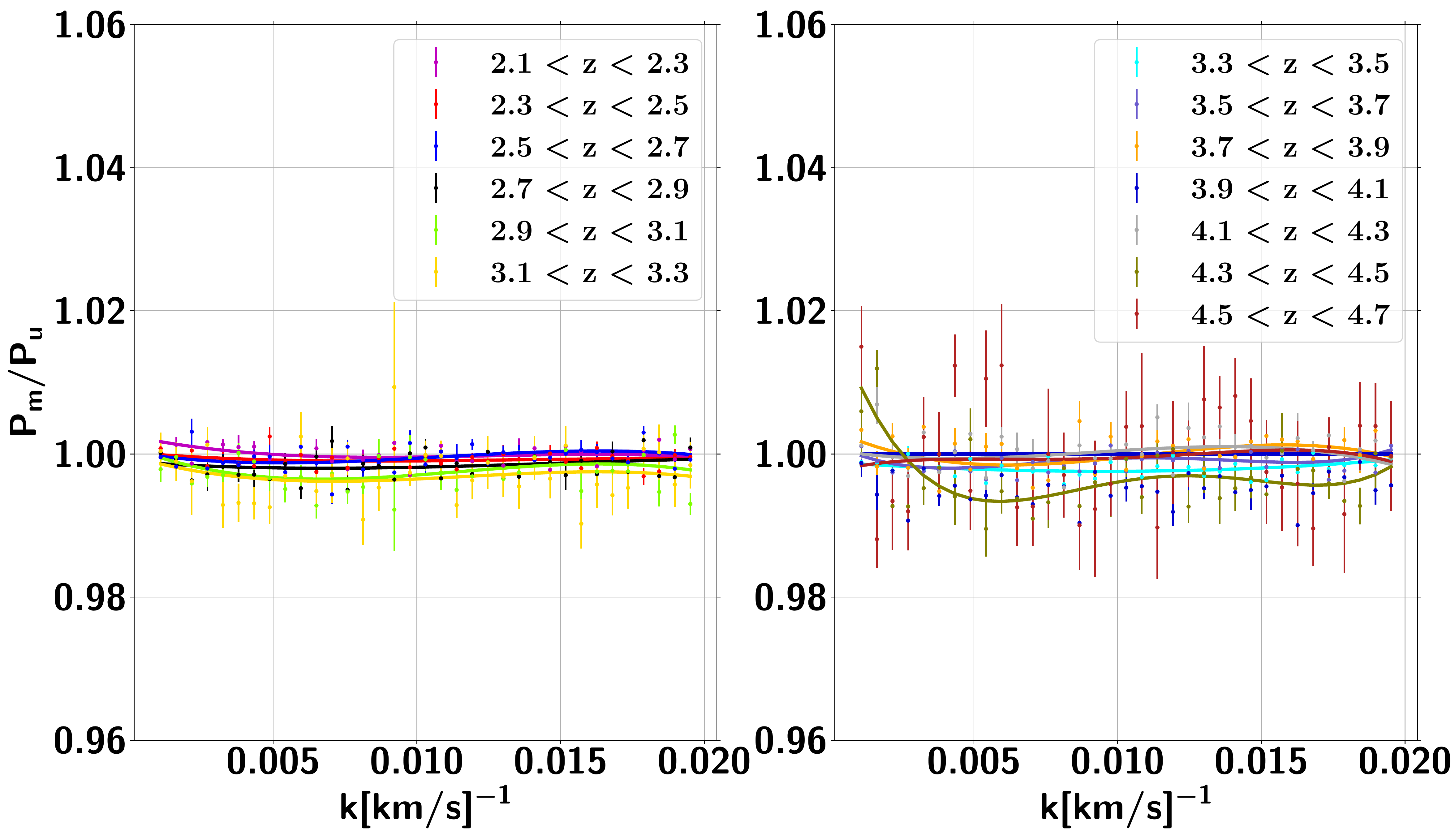,width = .97\textwidth}
\caption{\it Ratio of  the masked power spectrum $P_{m}$ to the unmasked power spectrum $P_{u}$, due to the masking of DLAs,  for all thirteen redshift bins from $z_{\rm Ly\alpha}=2.2$ to 4.6.  The bias in 2.1 $< z_{\rm Ly\alpha} <$ 4.3 and 4.3 $<z_{\rm Ly\alpha}<$ 4.7 is modeled by a $3^{rd}$ and a $4^{th}$-degree polynomial, respectively. } 
\label{fig:DLAbias}
\end{center}
\end{figure}

The impact of the sky emission line masking  is illustrated in Fig.~\ref{fig:SklBias}.  No strong sky line enters the forest for 2.7 $< z_{\rm Ly\alpha}<$ 3.3, which explains why no bias is observed in the corresponding redshift bins. The largest bias occurs for the $z_{\rm Ly\alpha}=4.2$ redshift bin, since it is the one with the largest number of sky lines  in the forest. For most of the impacted bins, we observe an underestimation on large scales  and an overestimation on small scales.  At first glance, the result is surprising, as we would naively expect masking to yield a loss of power. However, we can model the effect of pixel masking   as the convolution of the unmasked power spectrum by the squared Fourier Transform of the masking function. The masking function being either zero or one, according to whether the pixel is  masked or not,  it can be expressed as a sum of rectangular functions. As our initial power spectrum is decreasing with $k$, it appears natural to observe an excess of power on  large $k$-modes (i.e., small-scales).

The impact of the DLA masking  is presented in Fig.~\ref{fig:DLAbias}. As DLAs  are arbitrarily distributed in wavelength and in strength, and furthermore impact only a fraction of the forests at a given redshift, their masking induces a significantly smaller effect than that of sky lines. The low statistics of DLAs is responsible for the  scatter, in particular at high redshift.
The measured bias varies with redshift along with the fraction of  forests, in our selection, affected by DLAs. The fraction is less than 1\% at low redshift, around 5\% at intermediate redshifts and of 15\% at $z=4.6$.  The redshift evolution is in agreement with measurements in hydrodynamical simulations by \cite{Rogers2018} for instance, where their Table~1 shows an increasingly large fraction of their \lya\ spectra contaminated by DLAs as the redshift increases. This result is true for all categories, from small Lyman limit systems  to strong DLA absorbers. The trend is explained by the increasing total cross-section of DLAs with redshift.

Table~\ref{tab:bias} summarizes the sources of bias identified in the analysis. The final power spectra are corrected by the corresponding $k$- and $z$-dependent correction functions. The related systematic uncertainties associated to each of these corrections are discussed in the next section.
\begin{table}[htbp]
\caption{\it Maximum range of the corrections introduced at different steps of the analysis}
\begin{center}
\begin{tabular}{lc}
\hline
\hline
QSO continuum & 1.02 to 1.05\\
Masking of sky lines & 0.82 to 1.10 \\
Masking of DLAs & 0.99 to 1.00 \\
\hline
\end{tabular}
\end{center}
\label{tab:bias}
\end{table}


\section{Systematic uncertainties}
\label{sec:systs}
\begin{figure}[htbp]
\begin{center}
\epsfig{figure= 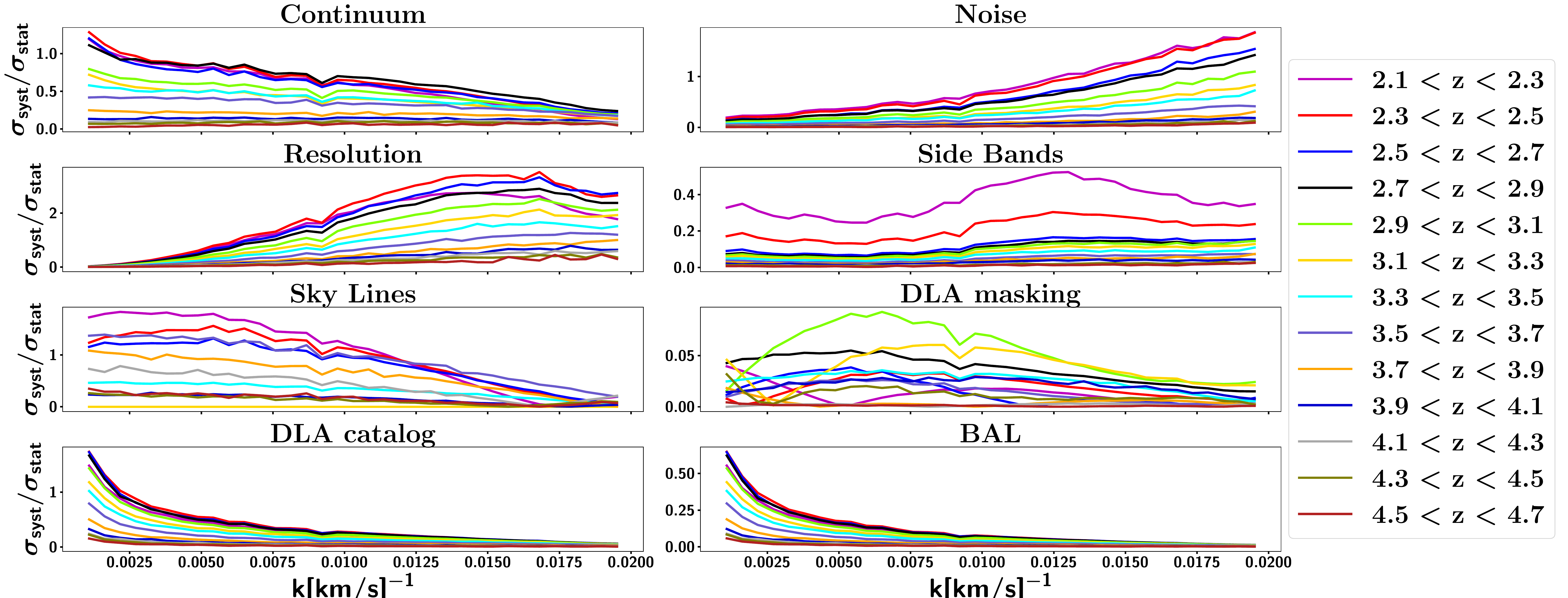, width = 16cm}
\caption{\it Ratios of the systematic to the statistic uncertainties, as a function of redshift and wave number, for the eight identified sources. From left to right and top to bottom are illustrated the  uncertainty ratios from the continuum estimation, the noise level, the spectral resolution, the side bands power spectra, the sky lines masking, the DLA masking, the DLA residual effects and the BAL features.} 
\label{fig:syst_stats}
\end{center}
\vspace{1cm}
\begin{center}
\epsfig{figure= 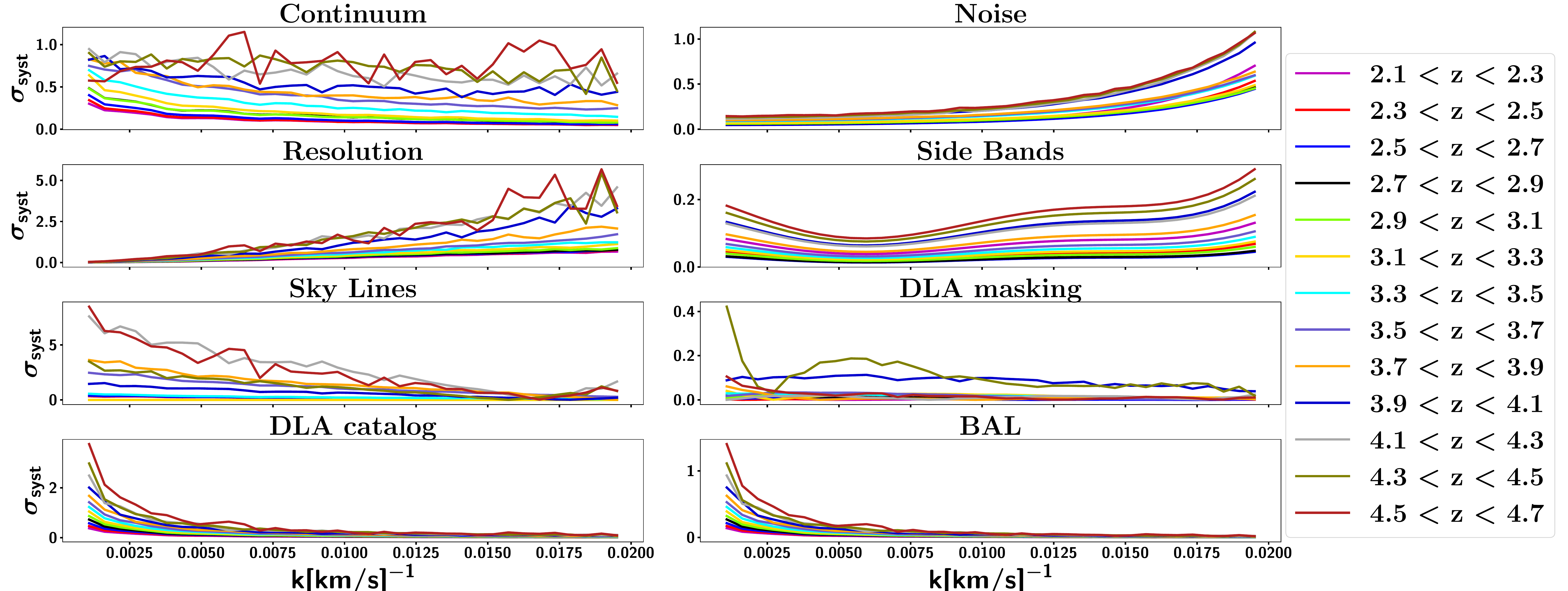, width = 16cm}
\caption{\it Absolute values of the systematic uncertainties, in ${\rm km/s}$, as a function of redshift and wave number, for the eight identified sources, in the same format as Fig.~\ref{fig:syst_stats}.} 
\label{fig:syst_abs}
\end{center}
\end{figure}

As we explained in the previous two sections, going from Eq.~\ref{eq:P1D_raw} (how $P^{\rm raw}$ is derived from observational quantities) to the final measurement of $P^{{\rm Ly}\alpha}$ requires selections  and power spectrum corrections at several stages of the analysis. These corrections and the impact of the selections are  each determined with their own degree of precision,  from which we infer a $k$- and $z$-dependent systematic uncertainty on the measurement of $P^{{\rm Ly}\alpha}$.  We identify eight systematic uncertainties:
\begin{itemize}
\item  Measurement of the quasar spectrum continuum
\item  Measurement of the quasar spectrum noise level
\item  Measurement of the spectrograph spectral resolution 
\item  Measurement of the power spectrum in side bands
\item  Effect of masking of the sky emission lines
\item  Effect of masking of the DLA absorbers
\item  Effect of the completeness of the DLA catalog
\item  Effect of the completeness of the BAL catalog
\end{itemize}
We now briefly describe  each of these systematic uncertainties. Their impact  is summarized in Figs.~\ref{fig:syst_stats} and~\ref{fig:syst_abs}.  The  systematic uncertainties we have identified are expected to be uncorrelated and can therefore be added in quadrature. We  give the values of each of them individually (in online fits files with the format of Table~\ref{tab:results}) so that other treatments can also be applied. 

\vskip 4pt
As explained in Sec. \ref{sec:cont}, the correction related to the determination of the continuum   is validated by assessing that consistent power spectra are obtained after application of the correction for the continuum estimate,  whichever form is used for the quasar-dependent term of the  continuum function. The agreement is at the 1\% level. We assign a 30\% uncertainty on the  correction measured, which leads to an uncertainty of comparable magnitude to the aforementioned agreement. 
This 30\% comes from the following assumption. As a conservative choice on  unknowns that could affect the value of the correction, we consider a shift  randomly ranging between ‘no correction' and ‘100\% of the correction’, which we describe by a uniform distribution between 0 and 1. The standard deviation of the distribution,  equal to $1/\sqrt{12}\sim 0.30$, quantifies the spread among the possible values, leading to a  systematic uncertainty  equal to 30\% of the correction. 
\vskip 4pt
The quasar spectrum noise level is determined through the procedure described in Sec.~\ref{sec:noise}. A ratio $\beta$ different from 1 is an indication of a small discrepancy between the measured noise power spectrum  and the one present in the coadded raw  spectrum that the SDSS pipeline delivers. We assign a systematic uncertainty on the resulting noise power spectrum equal to the 30\% of largest (1- $\beta$) term, all redshifts considered. The maximum value is obtained for $z_{\rm Ly\alpha}=2.2$, where the noise dominates $P^{raw}$.  Since the noise has a white power spectrum, the spectrograph window function (see Eqs.~\ref{eq:P1D_raw} and \ref{eq:pk}) make the impact of this systematic most significant on large $k$s.
\vskip 4pt
One of the main systematic uncertainties of this analysis is the knowledge of the spectral resolution $R$ which enters the window function term, $W^2(k,R,\Delta v)$. The spectral resolution is measured in SDSS  using arc lamps.  From several studies (see~\cite{Smee2013,Palanque-Delabrouille2013}), we derive that  the accuracy $\Delta R/R$ on the  measurement of $R$ is of order  $2\%$. To take  this uncertainty into account,  we compute the average resolution $\left< R\right>$ over the list of quasars that contribute to each redshift bin. The systematic uncertainty on $P_{1D}(k)$ is then given by $P_{1D}(k)\cdot (2k^2 R\Delta R)$. The quadratic $k$-term makes the large $k$-modes   more affected by this uncertainty. The mean resolution $\left< R\right>$ also varies with redshift, from  $81\,{\rm km/s}$  to $65\,{\rm km/s}$ with larger values for lower redshifts. This redshift-dependence induces a  larger impact for low-$z$ bins. 
\vskip 4pt
The power spectrum in the side bands, defined in Section~\ref{sec:SB}, is used to estimate the  power spectrum of the metal absorption and to correct for residual effects of the pipeline. However, the accuracy of these corrections are limited by the numbers of quasars with side bands in the relevant wavelength range. Therefore we propagate, as systematic uncertainties, the statistical errors on the determination of $P^{\rm SB_1}(k)$. As shown in Fig.~\ref{fig:SB1_zbins}, the shape of the power spectrum is obtained, for each redshift bin,  from the product of a universal sixth-degree polynomial derived from the average shape for all the quasars, and a first-degree polynomial in which both parameters are free. We vary the shapes according to the statistical errors on the latter parameters to estimate the systematic uncertainties. The  largest systematic uncertainties are obtained at high $k$  for the three redshift bins that contain the overlap between the two arms of the spectrograph: $ 5800 < \lambda< 6350 \,\AA$.
\vskip 4pt
The level of uncertainty on the correction of  sky line masking, computed by means of mock spectra, varies with the amplitude and shape of the  input power spectrum $P_g(k)$ and with the value of the  spectral resolution. These two input parameters were therefore chosen, in the mocks, to reproduce as well as possible the measured values of these inputs at all redshifts. The spectral resolution and input power spectrum were varied within observational limits; the measured variations of the derived corrections were at the level of 3\% and 5\%, respectively.  To include both dependences, we assign a conservative overall 30\% uncertainty  to  the correction for sky line masking. As shown in Sec.~\ref{sec:bmasking}, the largest effect occurs where the bias is largest, e.g.,  at  redshift  $z_{\rm Ly\alpha}=4.2$ for low $k$s.  In contrast, there are no sky line systematics for  redshifts  $2.7<z_{\rm Ly\alpha}<3.3$, since such forests contain no strong sky lines. 
\vskip 4pt
Like done in the masking of the sky lines, we assign a 30\% systematic uncertainty associated to the correction for the masking of DLA absorbers. Because the DLA masking yields at most a 1\% correction  (cf. Fig.~\ref{fig:DLAbias}), the related systematic is sub-dominant compared to all others. The DLA correction shows almost no dependence on $k$, and the redshift dependence is explained by the increasing percentage of contaminated forest. 
 \vskip 4pt
  \begin{figure}[htbp]
\begin{center}
\epsfig{figure= 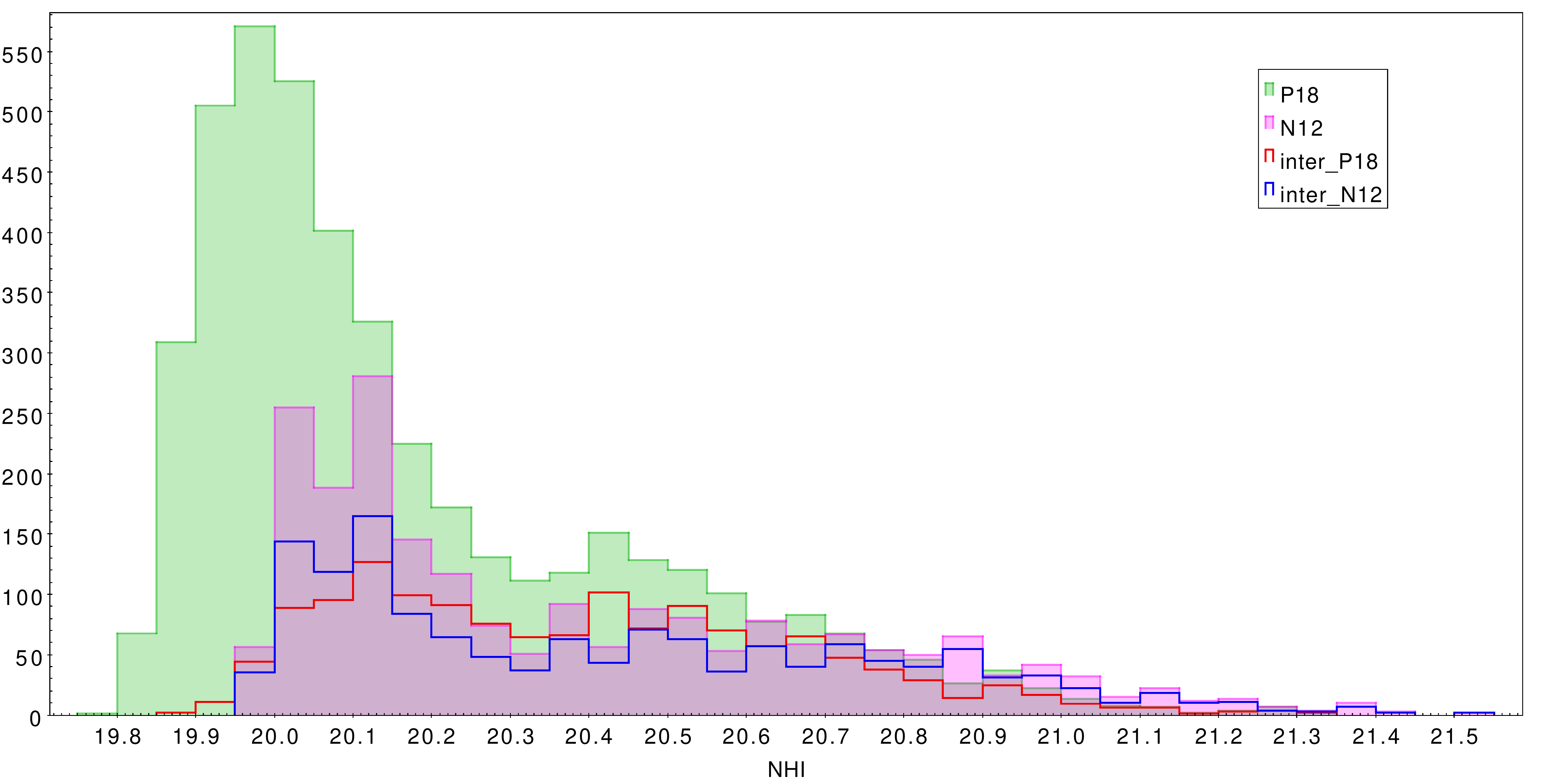,width = \textwidth}\\
\caption{\it Distribution of $\log N_{HI}/{\rm (cm^{-2})}$  for N12 in magenta and P18 in green. The solid lines show the $N_{HI}$ distribution of DLAs that are present in both catalogs. This illustrates the difference in the $N_{HI}$ estimation for common objects.} 
\label{fig:NHI_dist}
\vspace{1cm}
\epsfig{figure= 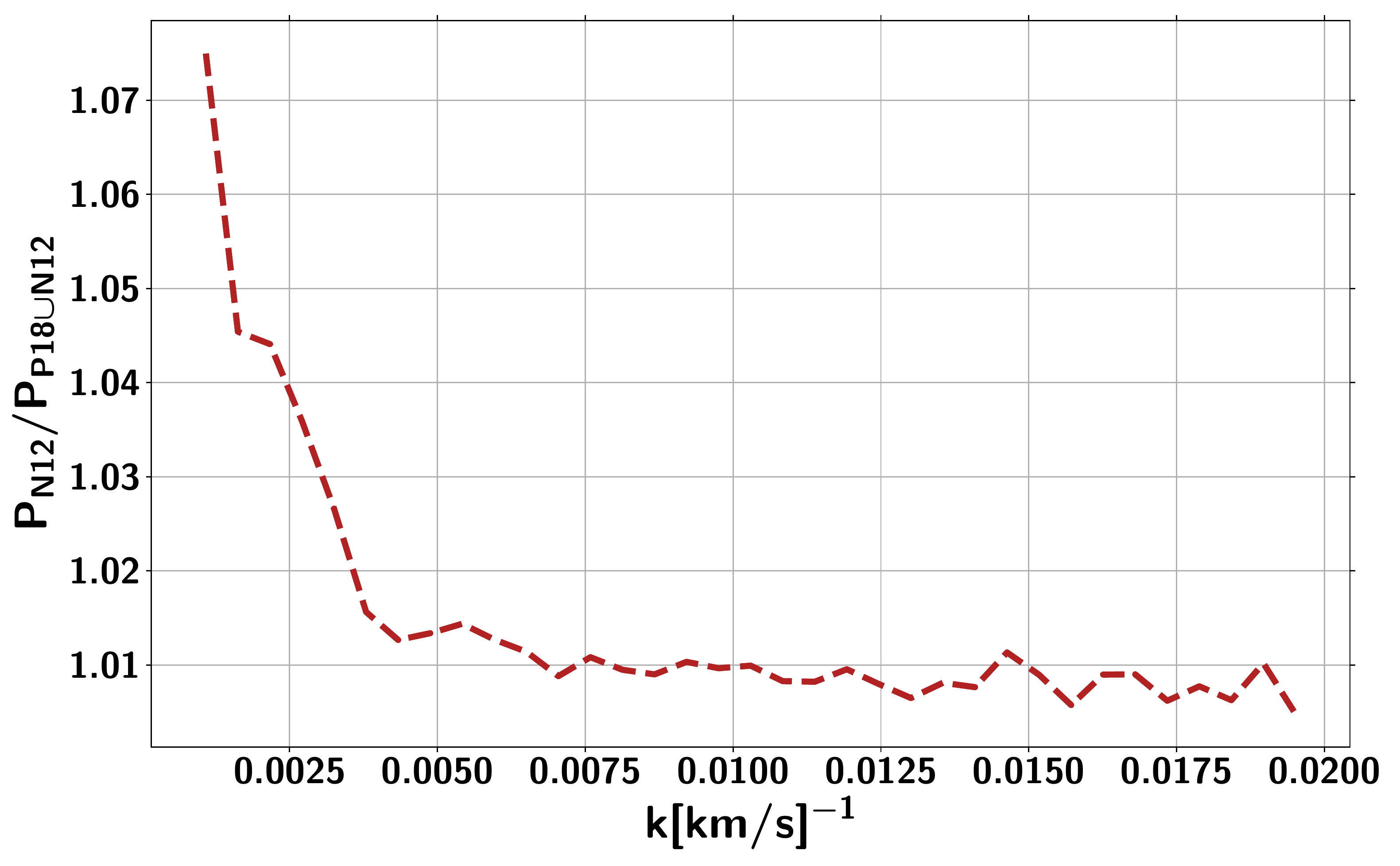,width = \textwidth}\\
\caption{\it Ratio of the power spectra using N12 to the union of P18 and N12 as DLA catalogs.  From the distribution shown in Fig.~\ref{fig:NHI_dist} it can be interpreted as the ratio of sub- and small- DLAs power spectra to the uncontamined power spectra.} 
\label{fig:subDLA_effect}
\end{center}
\end{figure}

The residual effect of unmasked DLAs was not taken into account in the systematics budget of PYB13. 
Our analysis uses the automated DLA catalog of N12~\cite{Noterdaeme2012}, as was done in~\cite{Bautista2017,DuMasDesBourboux2017}.
We compute a systematic uncertainty associated to this sample from the data themselves, using an alternative DLA catalog from P18~\cite{Parks2018}. 
Fig.~\ref{fig:NHI_dist} displays the distribution of the column density $N_{HI}$ for both catalogs. Both catalogs were optimized to identify DLAs, i.e., absorbers with $N_{HI} \ge 10^{20.3}\,{\rm cm^{-2}}$. P18 has a lower minimum column density and includes many more sub-DLAs and weak DLAs,   which still have a significant impact on the power spectra as demonstrated in~\cite{Rogers2018}.  P18 contains 4419 DLAs in the selected forest sample and N12 contains 2105 DLAs. 
We compare the resulting power spectrum to that obtained when masking with the superset of P18 {\em and} N12. When possible, we  select the  $N_{HI}$  from N12 to provide as consistent a comparison as possible.
 As shown in \cite{Rogers2018}, the impact of DLAs on the power spectrum strongly depends on the  $N_{HI}$ of the absorbers considered.
 Fig.~\ref{fig:subDLA_effect}  illustrates the  effect  of the additional DLAs of P18, integrated over all forest redshifts. Our results indicate a rise of a few percent on the largest scales ($k<0.003\,\rm (km/s)^{-1}$), in qualitative agreement with ~\cite{Rogers2018}. 
 We assign an  uncertainty equal to 30\% of  the ratio ($P_{N12}(k) / P_{P18 \cup N12}(k)$-1).
   \vskip 4pt
We reject from the analysis all quasars exhibiting BAL features, identified by a non-zero BI\_CIV flag. We consider that the automated procedure identifies the 80\% largest BALs with high efficiency, but could be incomplete for the 20\% faintest ones, i.e., those with ${\rm BI\_CIV}<170\,{\rm km/s}$. We thus compute the ratio $P_{{\rm BI\_CIV}>0} (k) / P_{{\rm BI\_CIV}>170}(k)$, and we assign a systematic uncertainty  equal to 100\% of this ratio. Unidentified BALs mostly affect the power spectrum on large scales, but the effect remains sub-dominant.


\section{Results}
\label{sec:results}
\subsection{Power spectrum}
\label{subsec:results_pk}
\begin{figure}[htbp]
\begin{center}
\epsfig{figure= 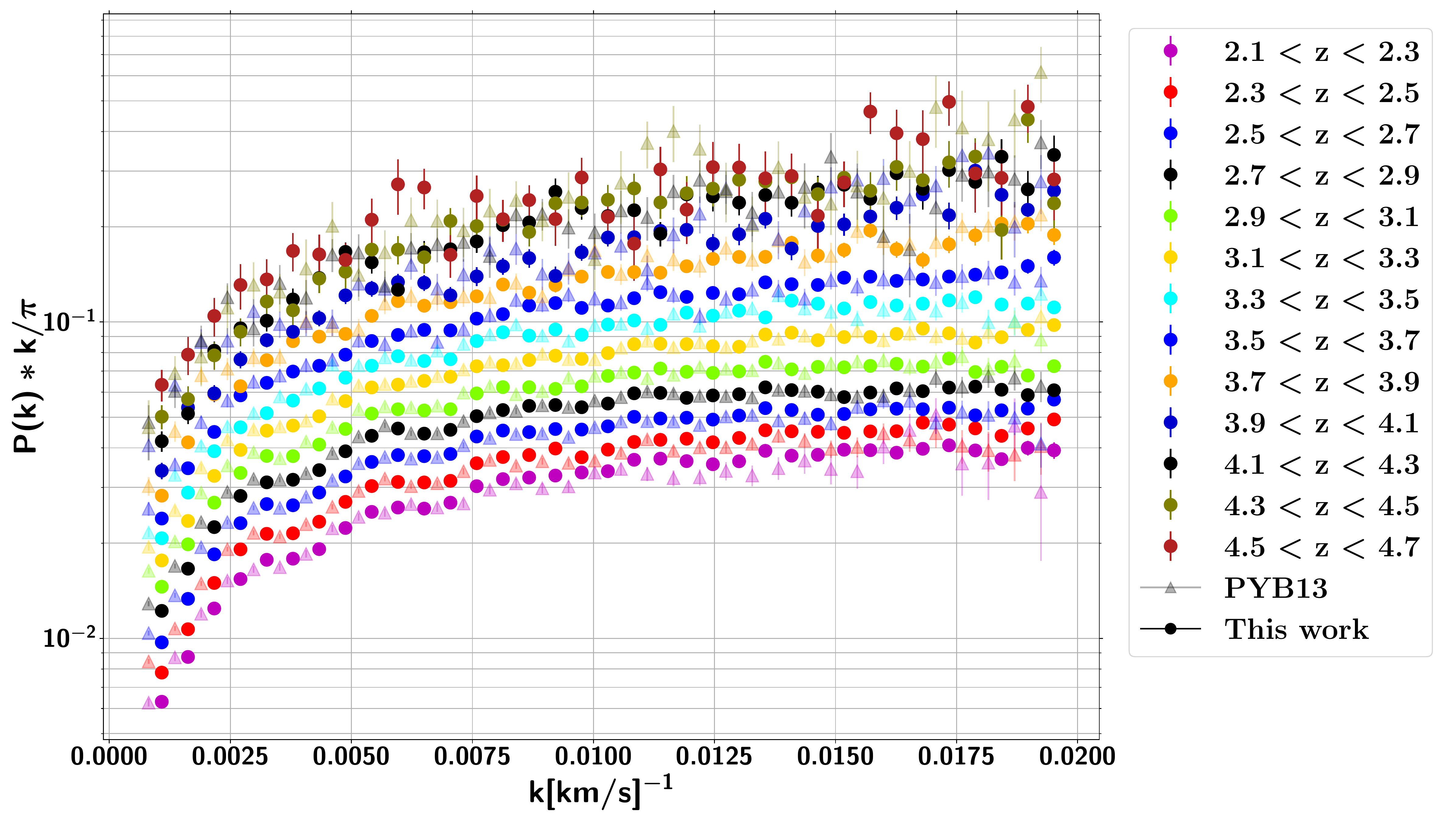,width = \linewidth}
\caption{\it  The 1D Ly$\alpha$ forest power spectrum. The measurements from PYB13 are shown in light colors, slightly shifted to smaller $k$'s for the better clarity. Error bars are statistical only. }
\label{fig:comp_DR9_eBOSS}
\end{center}
\end{figure}

Using the procedure described in the previous sections, we compute the 1D power spectrum over 13 redshift bins from $z_{\rm Ly\alpha}=2.2$ to 4.6, and over 35 modes from $k=10^{-3}$ to $k=0.02\, \rm (km/s)^{-1}$. The resulting power spectra are presented in Fig.~\ref{fig:comp_DR9_eBOSS}. The results are in excellent agreement with the  one published in PYB13, with no significant shift on any of the points, as also visible in the same figure. The errors are estimated in two ways, either from the rms of the distribution of the values of the power spectrum for a given $k$ and $z_{\rm Ly\alpha}$ over all contributing forests, or using a bootstrap approach. Both methods yield similar results.

The statistical uncertainties $\sigma_{\rm stat}$ are reduced by about a factor of two relative to PYB13 at all redshifts, due to the approximately four-fold increase of the selected quasar sample. The systematic uncertainties $\sigma_{\rm syst}$, in contrast, are increased by a about a factor of two, due to a more thorough investigation  of the possible sources of systematics that affect the measured power spectrum. The major difference arises from the study of the impact of the possible incompleteness of the BAL and DLA catalogs. Although the contribution of the resulting uncertainty is larger for large redshifts where $P_{1D}$ is larger  (cf. Fig.~\ref{fig:syst_abs}), the remarkably small value of the statistical uncertainty at small redshift makes the relative contribution of $\sigma_{\rm syst}$ more important at low redshift (cf. Fig.~\ref{fig:syst_stats}).
We also measure a $k$-dependent  bias, and hence a $k$-dependent systematic uncertainty, resulting from the procedure used to determine the quasar continuum. This feature was not observed in the previous method, which, however, was less sophisticated and did not include a quasar-dependent term.  
The systematic uncertainty related to the correction for the noise power is slightly reduced compared to PYB13. Other error contributions, such as systematics related to side bands, sky line and DLA masking, are similar to what was measured before.  

The agreement between the two analyses can be examined with a pull distribution, i.e.  the distribution of the difference in $P(k,z)$ divided by the statistical error on that difference. 
Because a  fraction of the quasars selected in the DR9 analysis of PYB13 are included in the present work, the combined statistical error  overestimates the error on the difference. However, given the  increased size of the data set and more optimized selection of the present analysis  (leading to a factor of two reduction of $\sigma_{\rm stat}$),  the overestimate is 10\% at most.
The pull distribution has a mean of $0.04\pm 0.05$, thus indicating excellent agreement on average, and a standard deviation of $1.18\pm0.10$. The spread slightly exceeds the one expected from  purely statistical effects. This result is  to be expected, since some steps of the analysis procedure  are affected by  different systematic uncertainties. To obtain a qualitative insight of the impact of  these systematics, a reasonable compromise is to add in quadrature the  systematic uncertainties from the present analysis only: we  include  sources of biases between the two pipelines while not double-counting the contributions that are common to both. This approach reduces the standard deviation of the ratio to 1.00, indicating that the differences between the power spectra resulting from PYB13 and  the present analysis are well explained by their  uncertainties. 

\begin{figure}[htbp]
\begin{center}
\epsfig{figure= 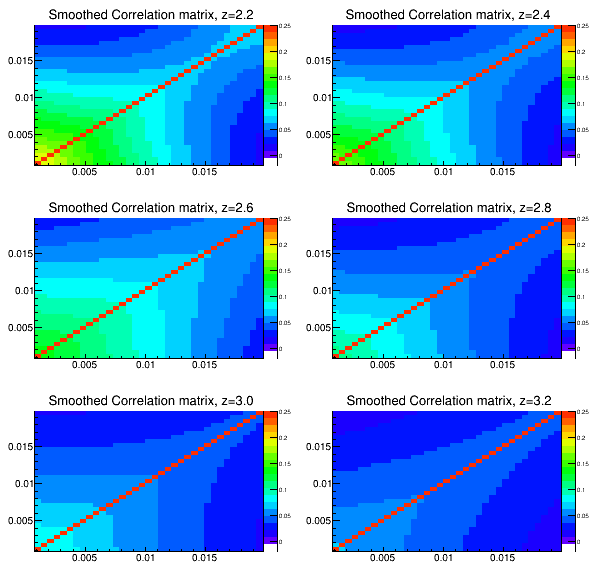,width = .9\linewidth}
\caption{\it  Correlation matrices between  $k$-modes for the redshift bins from $z=2.2$ to  $z=3.2$. Axes are $k$ modes in $\rm (km/s)^{-1}$. The color range is chosen to saturate at a correlation of 25\%. }
\label{fig:pk_cor1}
\end{center}
\end{figure}

\begin{figure}[htbp]
\begin{center}
\epsfig{figure= 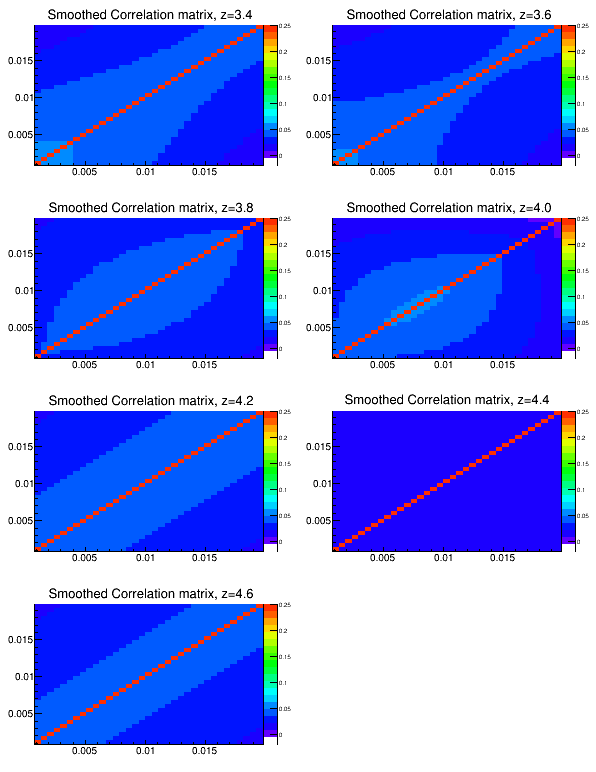,width = .9\linewidth}
\caption{\it  Correlation matrices between  $k$-modes for the redshift bins from $z=3.4$ to  $z=4.6$. Axes are $k$ modes in $\rm (km/s)^{-1}$. The color range is chosen to saturate at a correlation of 25\%.}
\label{fig:pk_cor2}
\end{center}
\end{figure}

\begin{figure}[htbp]
\begin{center}
\epsfig{figure= 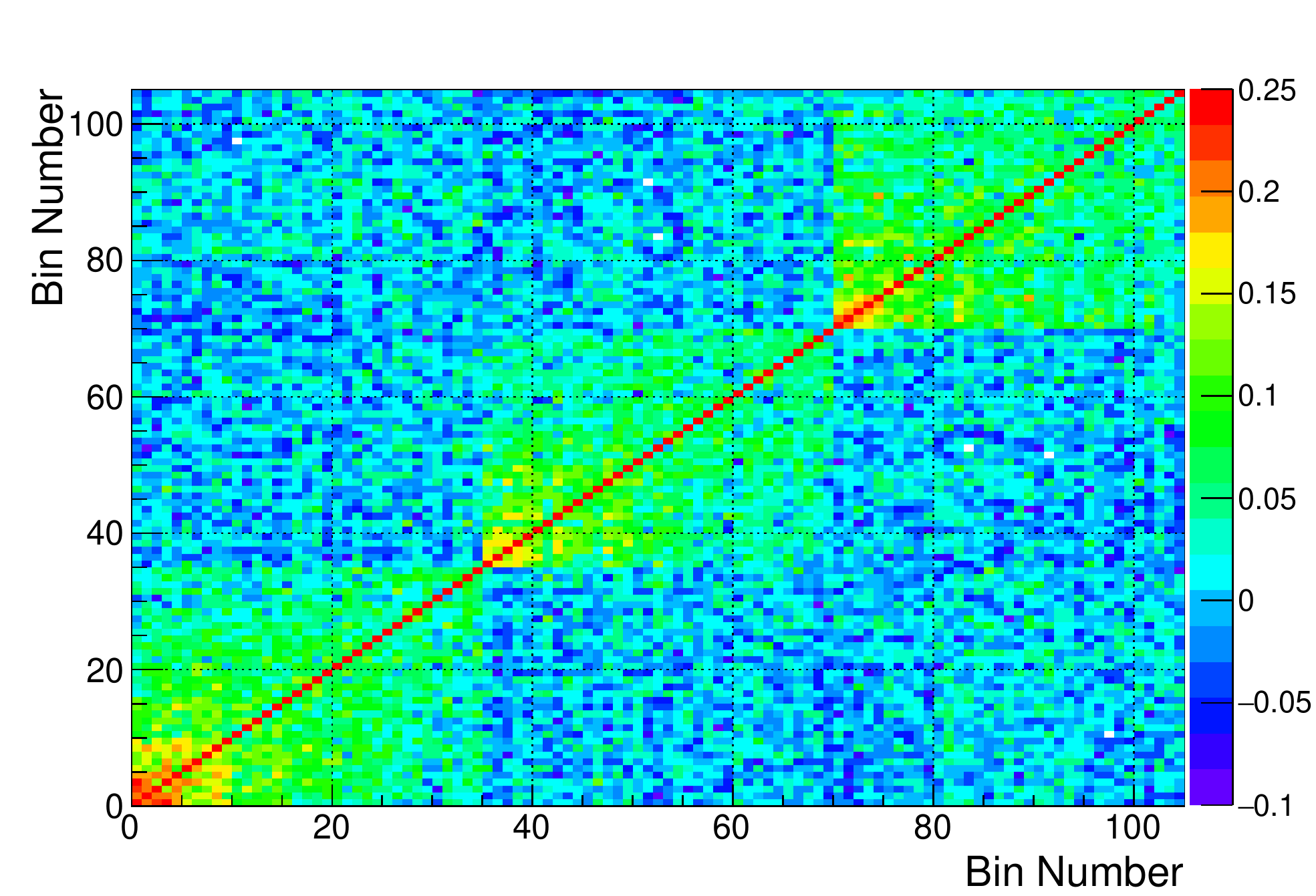,width = .9\linewidth}
\caption{\it  Correlation matrix between  $k$-modes for the  $z=2.2$,  $2.4$  and $2.6$ redshift bins. For each redshift, the power spectrum is measured over 35 $k$ bins. The axes of this correlation matrix therefore  have $3\times35$ k bins.  The diagonal  $35\times35$  sub-matrices correspond to the first three matrices of Fig.~\ref{fig:pk_cor1}. The color range is chosen to saturate at a correlation of 25\%. }
\label{fig:cor_mat_3x3}
\end{center}
\end{figure}

Finally,  Figs.~\ref{fig:pk_cor1} and \ref{fig:pk_cor2} display the correlation matrices measured for each of the 13 redshift bins, smoothed by second-order polynomials both along and across lines parallel to the diagonal elements of the matrix. The rms and the bootstrap approaches yield similar results. The correlation coefficients are 15 to 20\% at most at low redshift ($z_{\rm Ly\alpha}<3.0$) and on large scales ($k<0.01\,\rm (km/s)^{-1}$), and quickly decrease to values below 5\% otherwise. There is negligible correlation between redshift bins because the forest range (cf.  Sec.~\ref{sec:transmittedflux} for the wavelength coverage) 
has a redshift extension $\Delta z = 0.2$ at most. Each forest thus contributes to a single redshift bin. Moreover, the three sub-forests of a given quasar are processed independently to avoid induced correlations between them.  We check this assumption with a  bootstrap approach.  Figs.~\ref{fig:cor_mat_3x3} shows the result for the  $z=2.2,\,2.4$, and 2.6 redshift bins that present the strongest correlation on large scales. As expected, distinct redshift bins do not exhibit any correlation.

Table~\ref{tab:results} provides an extract of the measured power spectrum, as well as  statistical and  systematic uncertainties. The full table and correlation matrices are available online as fits files in the accompanying material attached to the paper.


\begin{table}

\caption{\it  Measured power spectrum $P^{Ly\alpha}$ in $km/s$ for each redshift bin $z_{\rm Ly\alpha}$ and scale $k$ in $\rm (km/s)^{-1}$. Also listed are the statistical uncertainty $\sigma_{\rm stat}$, the noise power $P^{noise}$, the side-band power $P^{\rm SB}$, and  each of the systematic uncertainties from the estimate of  (1)  continuum, (2) noise power, (3) spectrograph resolution, (4) side band power, (5) sky line masking, (6) DLA masking, (7) DLA catalog completeness, (8) BAL catalog completeness. Uncertainties, noise and side-band  power are  in $\rm km/s$.}  
\begin{center}
\begin{tabular}{cccccc}
\hline\hline
$z_{\rm Ly\alpha}$ & $k$ & $P^{Ly\alpha}$ & $\sigma_{\rm stat}$ & $P^{noise}$& $P^{\rm SB}$ \\
2.2 & 0.00108 & 19.2561 & 0.2527 & 2.5551 & 3.0573  \\
2.2 & 0.00163 & 17.4875 & 0.2152 & 2.5764 & 2.5723  \\
 ... \\
\hline
\end{tabular}
\begin{tabular}{cccccccc}
\hline\hline
$\sigma_{\rm sys\,1}$& $\sigma_{\rm sys\,2}$& $\sigma_{\rm sys\,3}$& $\sigma_{\rm sys\,4}$& $ \sigma_{\rm sys\,5}$& $\sigma_{\rm sys\,6}$& $\sigma_{\rm sys\,7}$& $\sigma_{\rm sys\,8}$\\
0.3008 & 0.0491 & 0.0060 & 0.0828 & 0.4361 & 0.0100 & 0.3946 & 0.1474 \\
0.2238 & 0.0495 & 0.0122 & 0.0752 & 0.3859 & 0.0075 & 0.2441 & 0.0893 \\
...  \\
\hline
\end{tabular}
\end{center}
\label{tab:results}
\end{table}

\subsection{Cosmological impact}

As a validation of these new measurements of the Ly$\alpha$ 1D flux power spectrum, 
we compute the cosmological impact of the data and compare to previous results by applying the same methodology and model. This allows us to highlight the differences induced by the new data. 
We  refer extensively to the earlier papers that reported constraints on cosmological parameters and on the  mass of active neutrinos~\cite{Palanque2015a,Palanque-Delabrouille2015}  using Ly$\alpha$ data  of PYB13 from the SDSS-III/BOSS survey. For the sake of simplicity, we will henceforth refer to~\cite{Palanque-Delabrouille2015} as PY15.  In this paper, we focus on results in the framework of $\Lambda$CDM with massless neutrinos. We leave the investigation of further models with massive neutrinos or warm dark matter  for future work.
Tab.~\ref{tab:param} summarizes the definition of the most relevant symbols used below. 

\begin{table}[htbp]
\caption{Definition of astrophysical and cosmological parameters}
\begin{center}
\begin{tabular}{p{2.9cm}ll}
\hline
\hline
Parameter& Definition \\
\hline
$\delta=\rho /  \left\langle \rho \right\rangle$  \dotfill  & Normalized baryonic density $\rho$ of IGM\\ 
$T$ \dotfill & Temperature of IGM modeled by $T=T_0  \cdot  \delta^{\gamma-1}$ \\
$T_0$ \dotfill & Normalization temperature of IGM at $z=3$ \\
$\gamma$ \dotfill &  Logarithmic slope of  $\delta$ dependence of IGM temperature  at $z=3$ \\
$\eta^{T_0}$ \dotfill &  Logarithmic slope of  redshift dependence of $T_0$ (different for $z<$ or $>3$) \\
$\eta^{\gamma}$  \dotfill &  Logarithmic slope of  redshift dependence of $\gamma$ \\
$A^\tau$  \dotfill &  Effective optical depth of Ly$\alpha$ absorption  at $z=3$\\
$\eta^\tau$  \dotfill &  Logarithmic slope of  redshift dependence of $A^\tau$   \\
$f_ {\rm{Si\,III}}$  \dotfill &  Fraction of Si\,III absorption relative to Ly$\alpha$ absorption\\
$f_ {\rm{Si\,II}}$  \dotfill &  Fraction of Si\,II absorption relative to Ly$\alpha$ absorption\\
$\Omega_m$ \dotfill & Matter fraction today (compared to critical density) \\
$H_0$ \dotfill &  Expansion rate today in km s$^{-1}$ Mpc$^{-1}$ \\
$\sigma_8$ \dotfill & RMS matter fluctuation amplitude today in linear theory \\
$n_s$ \dotfill & Scalar spectral index \\
\hline
\end{tabular}
\end{center}
\label{tab:param}
\end{table}%

To predict the Ly$\alpha$ flux power spectrum, we use the set of simulations extensively described in PY15 and \cite{Borde2014}. The simulations were run using a  parallel tree smoothed particle hydrodynamics  (tree-SPH) code {\sc Gadget-3}, an updated version of the public code  {\sc Gadget-2} \cite{Springel2001,Springel2005}. The simulations were started at $z=30$, with initial transfer functions and power spectra computed with {\sc CAMB}~\citep{Lewis2000}, and initial particle displacements generated with second-order Lagrangian Perturbation Theory.  Two particle types were included: collisionless dark matter, and gas.  

The simulations were run with cosmological parameters  centered on the Planck 2013 best-fit values  \cite{PlanckCollaboration2013}. Using simulations where one or two parameters at a time are assigned off-centered values, the first and second-order derivatives of the Ly$\alpha$ flux power spectrum were calculated with respect to each parameter,  which were used to derive a second-order Taylor expansion of the predicted Ly$\alpha$ flux power spectrum. The cosmological parameters cover the range $H_0=67.5\pm 5~{\rm km\,s^{-1}\,Mpc^{-1}}$, $\Omega_M=0.31\pm0.05$, $n_s=0.96\pm0.05$,  $\sigma_8=0.83\pm0.05$. All the runs used for this work assumed massless neutrinos, and were started with initial conditions using the same random seed. Snapshots were produced at regular intervals in redshift from $z = 4.6$ to 2.2, with $\Delta z = 0.2$.

We  tested the influence of assumptions on the IGM astrophysics  by running simulations for central and offset values of  relevant parameters. The photo-ionization rate of each simulation was fixed by requiring the effective optical depth at each redshift to follow the empirical law $\tau_{\rm eff}(z) =  A^\tau  (1+z) ^{\eta^\tau}$, with 
$A^\tau=0.0025 \pm 0.0020$ and $\eta^\tau=3.7\pm0.4$~in agreement with~\citep{Meiksin2009}. This renormalization was done at the post-processing stage, as justified in \cite{Theuns2005}, allowing a test of  the impact of different scalings without running new simulations. The IGM thermal history was chosen to reproduce the redshift evolution of the temperature-density parameters $T_0(z)$ and $\gamma(z)$ measured by~\cite{Becker2011}, where $T = T_0 \Delta^{\gamma-1}$  with $\gamma(z=3)=1.3\pm 0.3$ and $T_0 (z=3) = 14000\pm 7000\,$K. We extracted  particle samples from each of the simulation snapshots to compute $T_0(z)$ and $\gamma(z)$.  We used the quick-Ly$\alpha$ option to convert gas particles with over-densities exceeding $10^3$ and temperature below $10^5$~K into stars. 
One hundred thousand lines of sight are drawn randomly through the simulation box to compute the 1D flux power spectrum. The resulting statistical uncertainty on the power spectrum is small compared to that from the data. The simulation cosmic variance is added in quadrature to the simulation uncertainty, as explained in PYB15. 

As in PY15,  the determination of the coverage intervals of the fit parameters  is based on the `classical' confidence level method originally defined by \citet{neyman1937}. We start with the $\chi^2$  between a given  model defined by  $n$  parameters $\Theta=(\theta_{1},\ldots,\theta_{n})$, and  data measurements $x$ with Gaussian experimental errors $\sigma_{x}$.  We first determine the minimum $\chi^2_0$ with  all parameters free. To set a confidence level (CL) on each parameter $\theta_i$, we then scan the variable $\theta_i$: for each fixed value of $\theta_i$, we minimize  $\chi^2(x,\sigma_{x};\Theta)$ but with $n-1$ free parameters. The $\chi^2$ difference between the new minimum and  $\chi^2_0$ allows us to compute the CL on $\theta_i$,
\begin{equation}
{\rm CL}(\theta_i) = 1-\int_{\Delta \chi^2(\theta_i)}^{\infty}  f_{\chi^2}(t;N_{dof}) dt, \,\,{\rm with}\,\,  f_{\chi^2}(t;N_{dof})=\frac{e^{-t/2}t^{N_{dof}/2 -  1}}{\sqrt{2^{N_{dof}}} \Gamma(N_{dof}/2)}
\label{Eq:CL}
\end{equation}
where $\Gamma$ is the Gamma function and the number of degrees of freedom $N_{dof}$
is equal to 1. This profiling method can be easily extended to two variables. In this case, the minimizations are
performed for $n-2$ free parameters and the confidence level ${\rm CL}(\theta_i,\theta_j)$ is
derived from Eq.~\ref{Eq:CL} with $N_{dof}=2$. With this approach, the correlations between the variables are naturally taken into account and the minimization fit can explore the entire phase space of the  parameters. We do not apply any constraint to the astrophysical and cosmological parameters that we fit, which is equivalent  to having a wide uniform prior in Bayesian analysis.

\label{page:IGMparms}
The parameters in the  total $\chi^2$ belong to three categories. The first category models a flat $\Lambda$CDM cosmology with free $H_0$, $\Omega_M$, $n_s$,  and $\sigma_8$. The second category is a simplified model of the IGM, in which we let free the parameters given in table~\ref{tab:param}, namely $T_0$, $\gamma$, $\eta^{T_0}(z<3)$, $\eta^{T_0}(z>3)$, $\eta^{\gamma}$, $A^\tau$, $\eta^\tau$, and two amplitudes for the correlated absorption of Ly$\alpha$ with \ion{Si}{iii}  or \ion{Si}{ii}.  While the hydrodynamical simulations are run with a thermal history that follows the redshift evolution of~\cite{Becker2011}, we allow for additional freedom in the fit by modeling $T_0(z)$ and $\gamma(z)$ using a single power law for $\gamma$ and a broken power law at $z=3$ for $T_0$, as explained in PY15. Finally, the last category groups all nuisance parameters that allow us to account for uncertainties or corrections related to noise in the data, spectrograph resolution,  bias from the splicing technique,  UV fluctuations in the IGM, residual contamination from unmasked DLA, supernova and AGN feedbacks. Details on the fit parameters and on the dependance with scale and redshift of the nuisance parameters can be found  in PY15.  

As the data statistical uncertainties improve, systematic uncertainties  also have to be looked into with great care. In the present work, we did a thorough revision of all data-related systematics (cf. Sec.~\ref{sec:systs}). Simulation uncertainties were studied in detail in \cite{Borde2014} and in  PYB15. Among these, there are two dominant sources of systematics, both modeled as nuisance  parameters and left free in the fit. The first one is due to the impact of AGN feedback. It is currently modeled following the study of \cite{Viel12}. Further refinement of this study requires extensive hydrodynamical simulations, and constitutes the scope of a dedicated work (Chabanier et al., in prep). The second one is related to the use of the splicing approach. Its impact was measured relatively to non-spliced simulations of equivalent resolution and box-size as the spliced ones. The fit uses the shape that was derived  from these studies, and leaves two parameters free to allow flexibility in shape and amplitude around this template. With the advent of new, less computationally-expensive simulation codes such as Nyx~\cite{Almgren2013}, it might be possible in the near future to  directly produce simulations with the appropriate characteristics.    Other physical effects can affect  the 1D Ly-$\alpha$ flux power spectrum. These include  patchy \ion{He}{ii} reionization, for instance, that would add scatter in the temperature-density relation around $z=3$~\cite{McQuinn2009}, or various assumptions on the hydrogen reionization. Their investigation is left for future work.

\begin{figure}[htbp]
\begin{center}
\epsfig{figure= 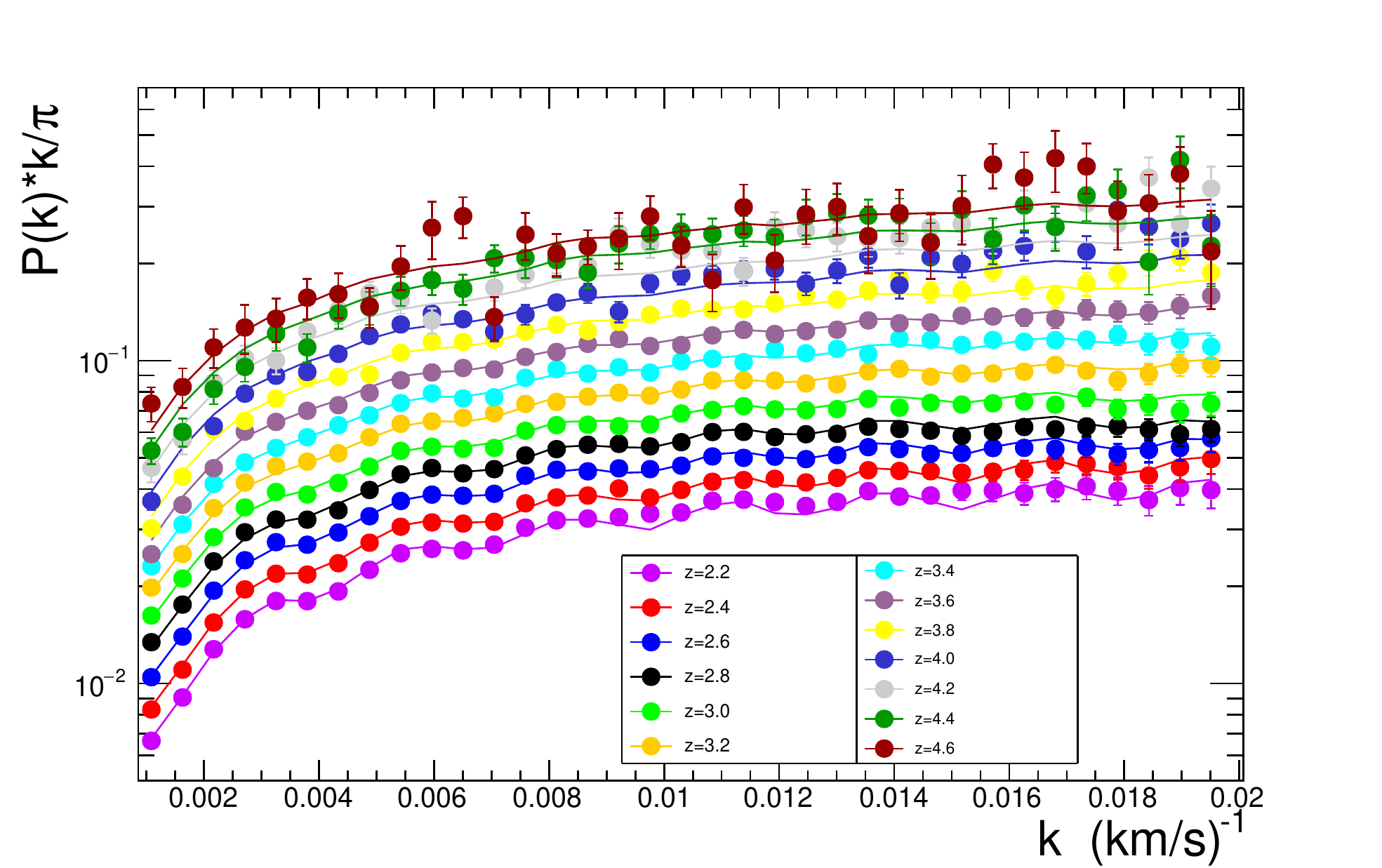,width = .9\linewidth}

\caption{\it 1D Ly$\alpha$ forest power spectrum for the analysis described in this paper.  Error bars include statistics and systematics added in quadrature. The solid curves show the best-fit model when considering Ly$\alpha$ data alone.  The oscillations  arise  from Ly$\alpha$-\ion{Si}{III} correlations, which occur at a  wavelength separation $\Delta \lambda = 9.2\,$\AA.} 
\label{fig:P1D_eBOSS_with_fit}
\end{center}
\end{figure}

In this cosmological analysis, we consider a $\Lambda$CDM cosmology with three types of massless neutrinos. The $\chi^2$ minimization yields the curves shown in Fig.~\ref{fig:P1D_eBOSS_with_fit} for the 13 redshift bins. 
The covariance matrices are computed using the correlation matrices described in Sec.~\ref{subsec:results_pk}, and the quadratic sum of the simulation uncertainty, the simulation cosmic variance, and the data statistical and systematic errors of  Table~\ref{tab:results}.
The agreement between  data and  fit is excellent, as the value $\chi^2= 435.8$ for  424 degrees of freedom,  corresponding to a p-value of 0.33, demonstrates. We check the robustness of the results by performing several data splits  shown in Table~\ref{tab:fit_results_split}.  The splits divide the data according to the spectral resolution, the SNR per pixel, the QSO catalogs (DR9, corresponding to ${\rm MJD}<55753$,   and post DR9), the spectrograph number (1 or 2), and the Galactic hemisphere (NGC, SGC). None of these data splits show any unexpected shift in the cosmological nor astrophysical parameters.

\begin{table}

\caption{\it Best-fit value and 68\% confidence levels of the cosmological parameters of the model fitted to the flux power spectrum. The dataset is split  in several subsamples based on the spectral resolution, the SNR per pixel,  the QSO catalog (DR9, post DR9), the spectrograph used and  the Galactic hemisphere (NGC, SGC).} 

{\small
\begin{center}
\begin{tabular}{lccccc}
\hline\hline
 Parameter &   Reference   &   $ \sigma_{\lambda} <80\,{\rm km\,s^{-1}}$ & ${\rm SNR}>4$   & ${\rm MJD}<55753$  &  ${\rm MJD}>55573$   \\

 \hline \\[-10pt]

$T_0$ (z=3) {\scriptsize($10^3$K)}                        & $10.3\pm1.9$           & $12.0\pm2.0$           & $11.7\pm1.9$           & $8.6\pm2.4$            & $11.4\pm1.9$ \\[2pt]
$\gamma$                                                               &  $0.8\pm0.1$            &  $0.8\pm0.1$            &  $0.9\pm0.1$            & $0.9\pm0.1$              & $0.8\pm0.1$\\ [2pt]
$\sigma_8$                                                              &  $0.820\pm0.021$    &  $0.826\pm0.022$    & $0.833 \pm 0.020$   & $0.850\pm 0.029$     & $ 0.819\pm 0.021$\\[2pt]
$n_s$                                                                       & $0.955\pm0.005$     & $0.957\pm0.006$     &  $0.951\pm0.008$    &  $0.945 \pm 0.007$   & $0.954 \pm 0.006$\\[2pt]
$\Omega_m$                                                           &  $0.269\pm0.009$     &  $0.270\pm0.010$    &  $0.276\pm0.012 $   &  $0.280\pm0.013$.    & $0.271 \pm 0.011$\\[2pt]
$H_0$~{\scriptsize(${\rm km~s^{-1}~Mpc^{-1}}$)}   & $67.1\pm1.0$           & $67.0\pm1.0$          &   $67.2 \pm 1.0$        & $67.3\pm 1.0$          & $67.0 \pm 1.0$ \\[2pt]
\hline

 &    Spectro $\#1$ & Spectro $\#2$  &  SGC & NGC  & \\
\cline{1-5}\\[-10pt]

$T_0$ (z=3) {\scriptsize($10^3$K)}                        & $10.3\pm1.9$           & $11.2\pm2.1$           & $11.3\pm3.1$            & $10.2\pm1.9$ & \\[2pt]
$\gamma$                                                               &  $0.9\pm0.1$            &  $0.8\pm0.1$            & $0.8\pm0.1$              & $0.8\pm0.1$ & \\ [2pt]
$\sigma_8$                                                              &  $0.826\pm0.023$    & $0.834 \pm 0.023$   & $0.794\pm 0.029$     & $ 0.825\pm 0.02$ \\[2pt]
$n_s$                                                                       & $0.963\pm0.006$     &  $0.939\pm0.007$    &  $0.960 \pm 0.011$   & $0.956 \pm 0.005$ \\[2pt]
$\Omega_m$                                                           &  $0.262\pm0.010$    &  $0.286\pm0.014 $   &  $0.263\pm0.013$.    & $0.271 \pm 0.010$  \\[2pt]
$H_0$~{\scriptsize(${\rm km~s^{-1}~Mpc^{-1}}$)}   & $66.9\pm1.0$          &   $67.3 \pm 1.0$        & $67.2\pm 1.0$          & $67.1 \pm 1.0$  \\[2pt]

\cline{1-5}

\end{tabular}
\end{center}
}

\label{tab:fit_results_split}
\end{table}

To study a potential effect due to residual DLAs not detected and thus not masked in the spectra, we introduce a multiplicative factor $1 - [1/(15000.0\;k-8.9) + 0.018]\cdot 0.2\cdot \alpha_{\rm DLA}$ to account for  a possible remaining contribution of high-density absorbers in the quasar spectra. This form is motivated by the study led by~\cite{McDonald2005}, and $\alpha_{\rm DLA}$ is free to vary in the fit. The fit converges to a null value of  $\alpha_{\rm DLA}$. Therefore in our standard fit procedure, we fix $\alpha_{\rm DLA}$ to zero.

\begin{figure}[htbp]
\begin{center}
\epsfig{figure= 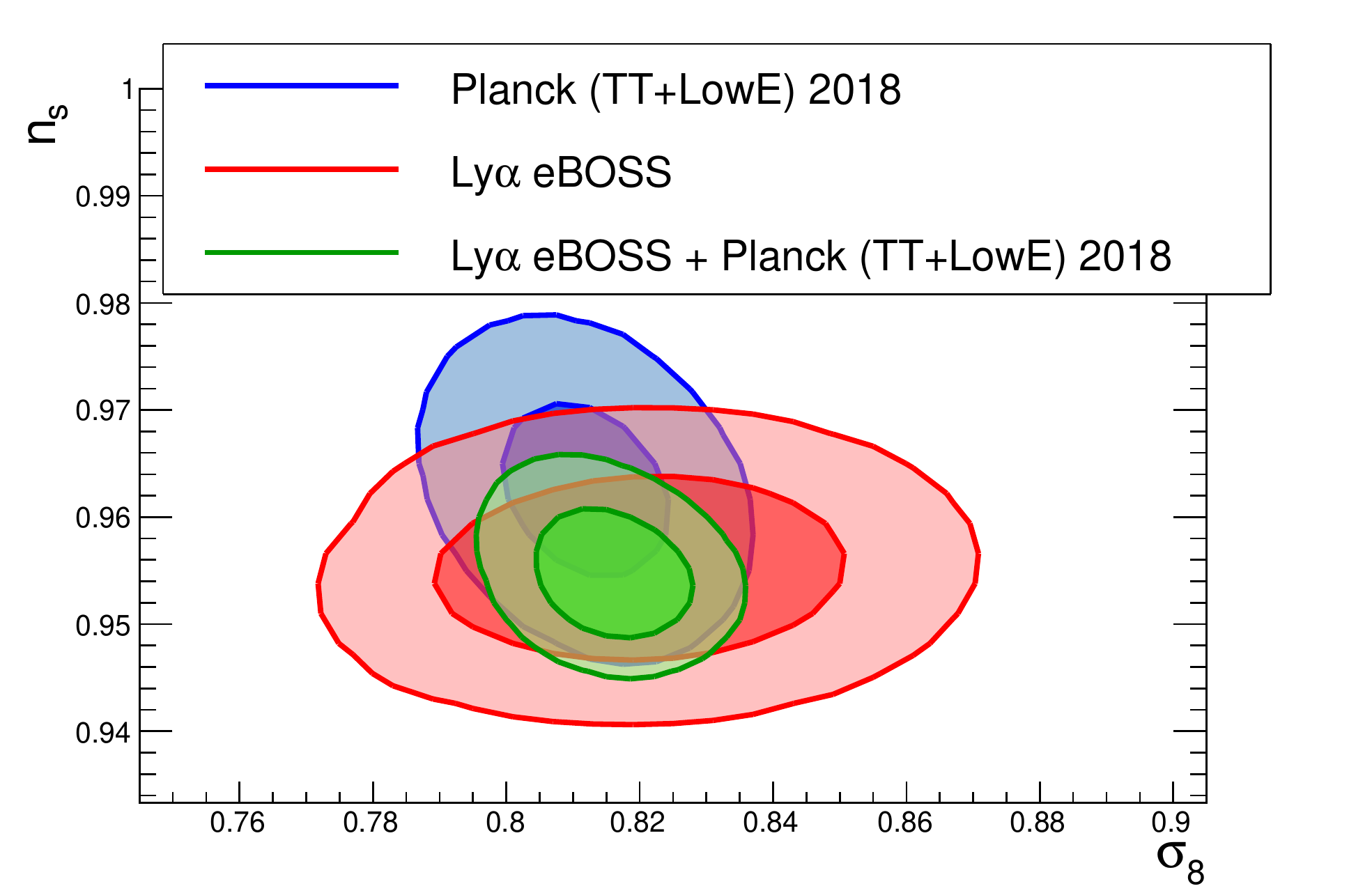,width = .9\linewidth}
\caption{\it   2D confidence level contours  in  $(\sigma_8 , n_s)$. The  68\% and 95\% confidence contours  are shown for  eBOSS Ly$\alpha$ data with a Gaussian constraint $H_0 = 67.3 \pm 1.0~{\rm km~s^{-1}~Mpc^{-1}}$ (red), for the Planck 2018  TT+lowE data (blue) and for the combination of  Ly$\alpha$ and Planck 2018 (green). The IGM thermal history  follows an over-simplified model  (cf. model description page \pageref{page:IGMparms}).}
\label{fig:sigma8_ns_Lya_Planck}
\vspace{1cm}
\epsfig{figure= 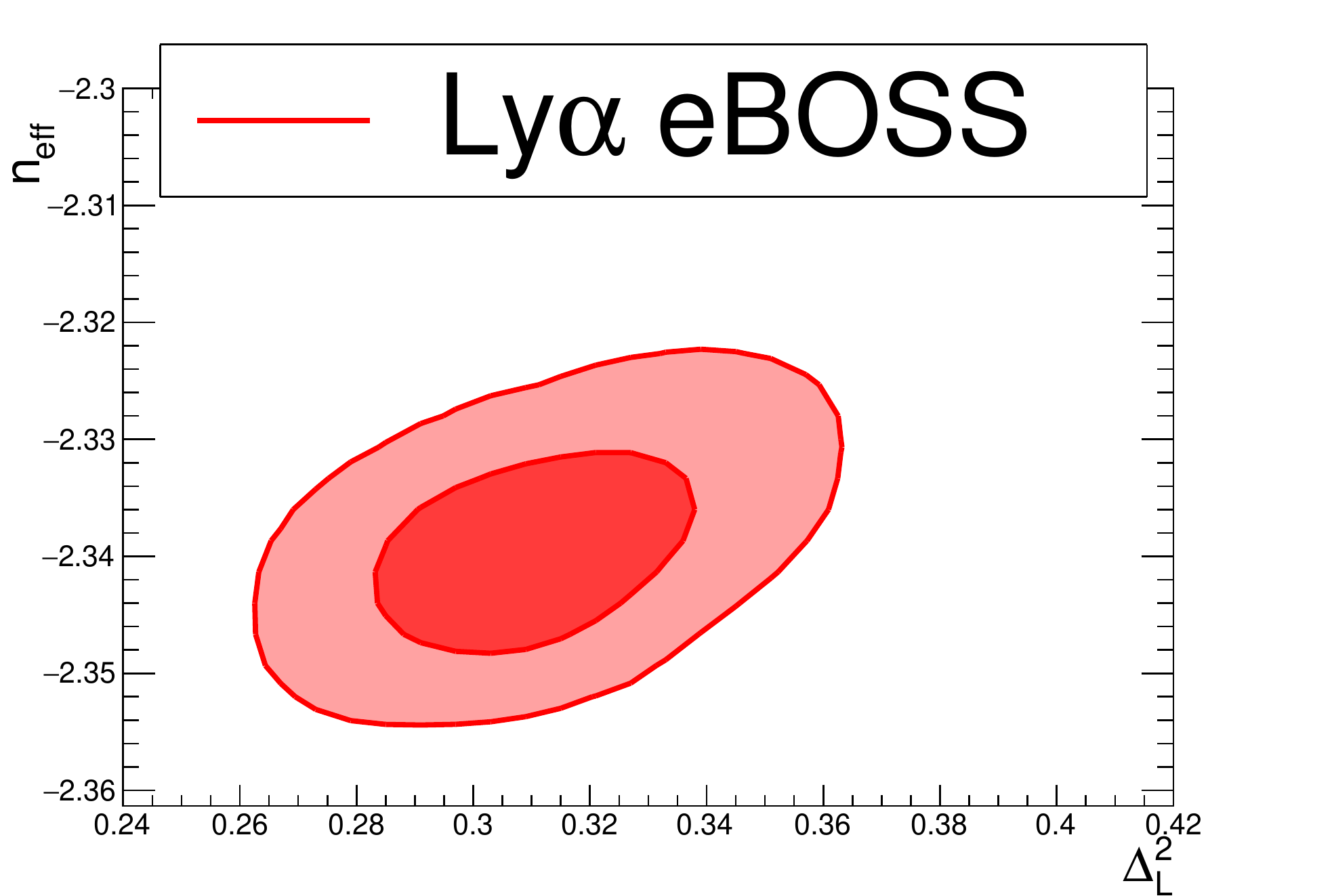,width = .9\linewidth}
\caption{\it   
Constraints on the effective slope $n_{\rm eff}$ and amplitude $\Delta^2_L$ of the linear power, measured at
$k_p=0.009 \,({\rm km/s})^{-1}$ and $z_p=3$ from Ly$\alpha$ data. The  68\% and 95\% confidence contours are obtained for  eBOSS Ly$\alpha$ data with a Gaussian constraint $H_0 = 67.3 \pm 1.0~{\rm km~s^{-1}~Mpc^{-1}}$. }
\label{fig:Delta2_neff_Lya}
\end{center}
\end{figure}

We summarize the cosmological results in Table~\ref{tab:fit_results} for different combinations (BOSS, eBOSS, XQ-100 and Planck). The results for Ly$\alpha$ alone are shown in columns (1) and  (2), respectively,  for BOSS DR9 (power spectrum computed in PYB13), and for eBOSS  (this work). The results are consistent for the two data sets. The only significant change is  for the scalar spectral index, $n_s$. The  mild tension on $n_s$ between Ly$\alpha$ and Planck in PY15 is no longer evident, as shown in Fig.~\ref{fig:sigma8_ns_Lya_Planck}. We perform a detailed, step-by-step comparison  of the two analyses  in order to understand  the shift on $n_s$. By reducing the eBOSS data set  to those in common with the DR9 quasar sample,  i.e., quasars observed before MJD=55753, we measure $n_s= 0.945 \pm 0.007$. We then replace the BAL and DLA automatic catalogs that we use for eBOSS by the visually identified DLA and BAL catalogs of DR9~\cite{Paris2012}. This change further shifts the spectral index  to  $n_s= 0.937 \pm 0.007$.  The latter value is fully compatible with the value published in PY15, $n_s= 0.937 \pm 0.009$. Therefore, the shift observed on the spectral index arises from the combination of a statistical fluctuation  and our usage of more complete  BAL and DLA catalogs. 

As  visible in Fig.~\ref{fig:P1D_eBOSS_with_fit},  the correlated absorptions by Ly$\alpha$ and \ion{Si}{III} are detected with high significance, and we measure $f_ {\rm{Si\,III}} = 6.0\,10^{-3} \pm  0.4 \,10^{-3}$. The Ly$\alpha$-\ion{Si}{II} correlated absorption is detected at 2~$\sigma$, with $f_ {\rm{Si\,II}} = 7.2\,10^{-4}  \pm 4.0\,10^{-4}$. 
 
In addition, we combine the  Ly$\alpha$ $\chi^2$ (imposing no constraint on $H_0$) with a $\chi^2$ term from  Planck 2018 data~\cite{Planck2018} that we derive from the central values and covariance matrices available in the official 2018 Planck repository. The results are given in column (3)  of Table~\ref{tab:fit_results}.  Fig.~\ref{fig:sigma8_ns_Lya_Planck} shows  the contours in the $n_s$-$\sigma_8$ plane for the Ly$\alpha$ data of column (2), for Planck data and for the combination  of column (3).  The best-fit value of  $n_s$ in  the combination of Ly$\alpha$ and Planck is slightly reduced compared to the one from Ly$\alpha$ or Planck alone, due to  the anti-correlation between $\Omega_m$ and $n_s$ in the Ly$\alpha$ data, and the higher value of $\Omega_m$ in Planck data. Finally, in column (4), we  add in the combination the Ly$\alpha$ power spectrum obtained from the higher resolution spectra observed with  the  VLT/XSHOOTER legacy survey (XQ- 100)~\cite{Yeche2017}.

\begin{table}

\caption{\it Best-fit value and 68\% confidence levels of the cosmological parameters of the model fitted to the flux power spectrum  
measured with the SDSS  or XQ-100 Ly$\alpha$  data~\cite{Yeche2017}  in  combination with  other data sets.  Column (1): results of PY15 with BOSS alone corresponding to PYB13.  Column (2): results with  eBOSS (this work). Column (3): results for the combined fit of  eBOSS (this work) and Planck 2018~\cite{Planck2018}.   Column (4): results for the combined fit of  eBOSS (this work), XQ-100 and Planck 2018. For columns (1-2), we used a Gaussian constraint, $H_0 = 67.3 \pm 1.0$. }  

\begin{center}
\begin{tabular}{lcccc}
\hline\hline
 &   Ly$\alpha$   &     Ly$\alpha$  &   Ly$\alpha$ &  Ly$\alpha$  \\
 &  BOSS & eBOSS & eBOSS  & eBOSS + XQ-100 \\
Parameter &  + $H_{0}^\mathrm{Gaussian}$  &  + $H_{0}^\mathrm{Gaussian}$  
&   + Planck &  + Planck  \\
 & (PY15)  & (This work)  & (TT+lowE) &   (TT+lowE) \\
 & (1) & (2)  &  (3) & (4) \\
 \hline \\[-10pt]

$T_0$ (z=3) {\scriptsize($10^3$K)}  & $8.9\pm3.9$   & $10.3\pm1.7$  & $11.3\pm1.6$  & $13.7\pm1.5$ \\[2pt]
$\gamma$   &  $0.9\pm0.2$   &  $0.8\pm0.1$     & $0.7\pm0.1$ & $0.9\pm0.1$\\ [2pt]
$\sigma_8$ &  $0.855\pm0.025$    & $0.820 \pm 0.021$     & $0.817\pm 0.007$ & $ 0.804\pm 0.008$\\[2pt]
$n_s$ &  $0.937\pm0.009$     &  $0.955\pm0.005$    &  $0.954 \pm 0.004$ & $0.961 \pm 0.004$\\[2pt]
$\Omega_m$  &  $0.288\pm0.012$     &  $0.269\pm0.009 $     &  $0.330\pm0.009$& $0.309 \pm 0.011$\\[2pt]
$H_0$~{\scriptsize(${\rm km~s^{-1}~Mpc^{-1}}$)}   & $67.1\pm1.0$      &   $67.1 \pm 1.0$    & $66.2\pm 0.6$ & $67.6 \pm 0.8$ \\[2pt]
\hline
\end{tabular}
\end{center}
\label{tab:fit_results}
\end{table}

A different parameterization has been used in some previously published results~\cite{McDonald2005}: the dimensionless amplitude $\Delta^2_L(k,z)\equiv k^3 P_L(k,z)/2\pi^2 $ and the logarithmic slope $n_{\rm eff}(k,z) \equiv d\ln P_L/d\ln k$ of the linear power spectrum $P_L$, both evaluated at a pivot redshift $z_p$ and pivot wave number $k_p$. For comparison with these previous results, we provide  $\Delta^2_L(k,z)$ and $n_{\rm eff}(k,z)$ for the present analysis. 
Figure~\ref{fig:Delta2_neff_Lya} illustrates our constraints using this parameterization, with $k_p=0.009 \,{\rm (km/s)^{-1}}$ and $z_p=3$, a central position in the medium-resolution SDSS Lyman-$\alpha$ data. The Ly$\alpha$ results  correspond to an amplitude  $\Delta^2_L = 0.31\pm 0.02$ and an effective slope $n_{\rm eff} = -2.339 \pm 0.006$.


\section{Conclusions}
\label{sec:conclusion}

Using the entirety of the BOSS and first-year eBOSS  data, this paper provides a measurement of the  Ly$\alpha$ forest 1D flux power spectrum covering a large redshift range from $z_{\rm Ly\alpha}=2.2$ to 4.6, and on scales ranging from $k=0.001$ to $0.020\:{\rm (km/s)^{-1}}$.   We perform a stringent selection of Ly$\alpha$ forest data. To ensure high confidence in the redshift assignment, for instance, we only select quasars that have been  visually inspected. We also reject low signal-to-noise ratio spectra to optimize the statistical precision of our results. From a parent sample of 180,413 quasars this tight selection leads to  43,751 quasars  used for the analysis presented in this paper.

The large quasar parent sample and optimized selection allow us to reduce the statistical uncertainty of the measurement by a factor of two compared to the previous  publication of the BOSS collaboration~\cite{Palanque-Delabrouille2013} referred to as PYB13. The increased data set also allows the addition of a new  redshift bin at  $z_{\rm Ly\alpha}=4.6$. Despite the lower statistical precision of this bin, it carries useful information from an earlier epoch in the history of the Universe, where the clustering is less affected by  non linearities  than at lower redshift. This redshift bin is therefore highly valuable  to constrain dark matter properties, for instance.

We performed an extensive thorough investigation of the systematic uncertainties affecting the measurement. To  study  systematic effects related to the analysis procedure, we employed mocks  tuned to match the level of the data power spectrum as well as the evolution of the spectrograph resolution with observed wavelength. We processed these mocks with the exact same pipeline as were the data. To estimate uncertainties related to the spectroscopic pipeline,  we used science data themselves, as well as calibration data. 
The major sources of uncertainty on small scales come from the precision on the determination of the spectrograph resolution, and from the estimation of the noise power. The latter largely dominates over the cosmological power in particular at low redshift. These two issues should be improved with the next generation WEAVE-QSO~\cite{Pieri2016} and Dark Energy Spectroscopic Instrument (DESI)~\cite{DESI2016} projects, which will have almost twice the spectral resolution and a higher signal-to-noise ratio for $z_{\rm qso}>2.1$ quasars. 
On large scales, the dominant source of uncertainty arises from the incompleteness of the BAL and DLA catalogs. The data quality in next generation surveys will improve their identification. Furthermore, work is on-going to   improve their automated detection.

We performed a detailed comparison of the results  obtained in this analysis with those of PYB13. The power spectra are in excellent agreement. Using the  model  of \cite{Palanque2015a,Palanque-Delabrouille2015} where the power spectrum is described by four cosmological parameters, five parameters for a simplified model of the IGM thermal history, two  for the mean transmitted flux, and nuisance parameters for astrophysical and instrumental artefacts, all free to vary in the fit, we obtain comparable cosmological results as previously published. The main change is on the scalar spectral index $n_s$. The best-fit value is now $0.955 \pm 0.005$, in agreement with the result found by the Planck collaboration for a flat $\Lambda$CDM model; prior to our work, there was a tension at the  $2\sigma$ level. We identified the causes of this increase on $n_s$: about half of the increase is produced by  a statistical fluctuation of the first-year subset of quasars, the other half is due to the change in the DLA catalog used to veto forests affected by DLA features, from a visual catalog in PYB13 to an automated  catalog in the present work. DLAs add  power on large scales, and the improved completeness of the new catalog reduces the power on such scales, thus increasing $n_s$.

In addition to the  measurement of the  Ly$\alpha$ forest 1D flux power spectrum, we also provide the statistical uncertainty and each of  the systematic uncertainties.  In subsequent analyses, we suggest  combining all systematic errors in quadrature. In cases where a motivated model can be used to account for the impact on the power spectrum of a given systematic, then we suggest omitting its contribution to the overall error budget and marginalizing over the model parameters in the fit.  We provide the full results of our analysis of the  Ly$\alpha$ forest 1D flux power spectrum as fits files in the accompanying material.


\acknowledgments

We acknowledge PRACE (Partnership for Advanced Computing in Europe) for awarding us access to resources Curie thin nodes and Curie fat nodes, based in France at TGCC, under allocation numbers 2010PA2777, 2014102371 and 2012071264. 
This work was also granted access to the resources of CCRT under the allocation 2013-t2013047004 made by
GENCI (Grand Equipement National de Calcul Intensif).
We thank Cl\'ement Besson for helpful contributions to the understanding of the estimate of the noise power spectrum. N.P.-D. and Ch.Y.  acknowledge  support from grant ANR-17-CE31-0024-01 of Agence Nationale de la Recherche for the NILAC project.\\
 Funding for the Sloan Digital Sky Survey IV has been provided by the Alfred P. Sloan Foundation, the U.S. Department of Energy Office of Science, and the Participating Institutions. SDSS-IV acknowledges support and resources from the Center for High-Performance computing at the University of Utah. The SDSS web site is www.sdss.org. SDSS-IV is managed by the Astrophysical Research Consortium for the Participating Institutions of the SDSS Collaboration including the Brazilian Participation Group, the Carnegie Institution for Science, Carnegie Mellon University, the Chilean Participation Group, the French Participation Group, Harvard-Smithsonian Center for Astrophysics, Instituto de Astrof\'isica de Canarias, The Johns Hopkins University, Kavli Institute for the Physics and Mathematics of the Universe (IPMU) / University of Tokyo, Lawrence Berkeley National Laboratory, Leibniz Institut f\"ur Astrophysik Potsdam (AIP), Max-Planck-Institut f\"ur Astronomie (MPIA Heidelberg), Max-Planck-Institut f\"ur Astrophysik (MPA Garching), Max-Planck-Institut f\"ur Extraterrestrische Physik (MPE), National Astronomical Observatories of China, New Mexico State University, New York University, University of Notre Dame, Observat\'ario Nacional / MCTI, The Ohio State University, Pennsylvania State University, Shanghai Astronomical Observatory, United Kingdom Participation Group, Universidad Nacional Aut\'onoma de M\'exico, University of Arizona, University of Colorado Boulder, University of Oxford, University of Portsmouth, University of Utah, University of Virginia, University of Washington, University of Wisconsin, Vanderbilt University, and Yale University.

\newpage
\bibliographystyle{unsrtnat_arxiv}
\bibliography{biblio}

\end{document}